\documentclass[oneside,11pt]{article}

\usepackage{times,url,mathrsfs}
\usepackage{amsmath,wrapfig,color,bigfoot}
\usepackage{amsfonts}
\usepackage{graphicx}
\usepackage{subfigure}
\newtheorem{theorem}{Theorem}
\newtheorem{lemma}{Lemma}

\begin{document}

\title{\huge Compressed Counting Meets Compressed Sensing\vspace{0.5in}}
\author{ \textbf{Ping Li} \vspace{0.05in}\\
         Department of Statistics \& Biostatistics\\
         Department of Computer Science\\
       Rutgers University\\
          Piscataway, NJ 08854\\
       \texttt{pingli@stat.rutgers.edu}\\\\\\
       \and
         \textbf{Cun-Hui Zhang} \vspace{0.05in}\\
         Department of Statistics \& Biostatistics\\
         Rutgers University\\
         Piscataway, NJ 08854\\
         \texttt{czhang@stat.rutgers.edu}
       \and
         \textbf{Tong Zhang}\vspace{0.05in}\\
         Department of Statistics \& Biostatistics\\
         Rutgers University\\
         Piscataway, NJ 08854\\
        \texttt{tongz@rci.rutgers.edu}
}

\date{}
\maketitle

\begin{abstract}

\noindent Compressed\footnote{The work was presented at Simons Institute Workshop on {\em Succinct Data Representations and Applications} in September  2013.} sensing (sparse signal recovery) has been a popular and important research topic in recent years. By observing that natural signals  are often nonnegative, we propose a new framework for nonnegative signal recovery using {\em Compressed Counting (CC)}. CC is a technique built on  {\em maximally-skewed $\alpha$-stable random projections} originally developed for data stream computations.  Our recovery procedure is computationally very  efficient in that it requires only one linear scan of the coordinates. \\

\noindent In our settings, the signal $\mathbf{x}\in\mathbb{R}^N$ is assumed to be nonnegative, i.e., $x_i\geq 0, \forall \ i$.  Our analysis demonstrates that, when $\alpha\in(0,\ 0.5]$, it suffices to use  $M=(C_\alpha+o(1)) \epsilon^{-\alpha} \left(\sum_{i=1}^N x_i^\alpha\right)\log N/\delta$ measurements so that, with probability $1-\delta$, all coordinates will be recovered within $\epsilon$ additive precision, in one scan of the coordinates. The constant $C_\alpha=1$ when $\alpha\rightarrow0$ and $C_\alpha=\pi/2$ when $\alpha=0.5$. In particular, when $\alpha\rightarrow0$, the required number of measurements is essentially $M=K\log N/\delta$, where $K = \sum_{i=1}^N 1\{x_i\neq 0\}$ is the number of nonzero coordinates of the signal.

\end{abstract}

\newpage\clearpage

\section{Introduction}

In this paper, we develop a new framework for \textbf{compressed sensing} (sparse signal recovery)~\cite{Article:Stark_89,Article:Huo_JIT01,Article:Cormode_05,Article:Donoho_CS_JIT06,Article:Candes_Robust_JIT06}. We focus on nonnegative sparse signals, i.e.,  $\mathbf{x}\in\mathbb{R}^{N}$ and $x_i\geq 0, \forall\ i$. Note that real-world signals are often nonnegative. We consider the scenario in which neither the magnitudes nor the locations of the nonzero entries of  $\mathbf{x}$ are unknown (e.g., data streams).  The task of compressed sensing is to recover  the locations and magnitudes of the nonzero entries. Our framework differs from mainstream  work in that we use maximally-skewed $\alpha$-stable distributions for generating our design matrix, while classical compressed sensing algorithms typically adopt Gaussian or Gaussian-like distributions (e.g., distributions with finite variances). The use of skewed stable random projections was originally developed in~\cite{Proc:Li_SODA09,Proc:Li_UAI09,Proc:Li_Zhang_COLT11}, named \textbf{Compressed Counting (CC)}, in the context of data stream computations. Note that in this paper we focus on dense design matrix and   leave the potential use of ``very sparse stable random projections''~\cite{Proc:Li_KDD07} for sparse recovery as future work, which will connect this line of work with the well-known ``sparse matrix'' algorithm~\cite{Article:Gilbert_IEEE10}.\\

In compressed sensing, the standard procedure first collects $M$ non-adaptive linear measurements
\begin{align}
y_j = \sum_{i=1}^N x_i s_{ij},\ \ j = 1, 2, ..., M
\end{align}
and then reconstructs the signal  $\mathbf{x}$ from  the measurements, $y_j$, and the design matrix, $s_{ij}$. In this context, the design matrix is indeed ``designed'' in that one can manually generate the entries to facilitate signal recovery. In fact, the design matrix can be integrated in the sensing hardware (e.g., cameras, scanners, or other sensors).  In classical settings, entries of the design matrix, $s_{ij}$, are typically sampled from Gaussian or Gaussian-like distributions. The recovery algorithms are often based on linear programming ({\em basis pursuit})~\cite{Article:Chen98} or greedy pursuit algorithms such  as {\em orthogonal matching pursuit}~\cite{Article:Mallat93,Proc:OMP_NIPS09,Article:Tropp_JIT04}. In general, LP is computationally  expensive. OMP might be  faster although it still requires scanning the coordinates $K$ times. \\

It would be  desirable to develop a new framework for sparse recovery which is much faster than linear programming decoding (and other algorithms) without requiring more measurements.  It would be also desirable if the method is robust against measurement noises and is applicable to \textbf{data streams}. In this paper, our method  meets these requirements by sampling $s_{ij}$ from  maximally-skewed $\alpha$-stable distributions~\cite{Book:Zolotarev_86}.

\subsection{Maximally-Skewed Stable Distributions}

In our proposal, we sample entries of the design matrix $s_{ij}$ from an $\alpha$-stable {maximally-skewed} distribution, denoted by $S(\alpha,1,1)$, where the first ``1'' denotes maximal skewness and the second ``1'' denotes unit scale.  If a random variable $Z\sim S(\alpha,1,1)$, then its characteristic function is
\begin{align}
{\mathscr{F}}_Z(\lambda) &= \text{E}\exp\left(\sqrt{-1}Z\lambda\right)
= \exp\left(-|\lambda|^\alpha\left(1-\text{sign}(\lambda)\sqrt{-1}\tan\left(\frac{\pi\alpha}{2}\right)\right)\right), \hspace{0.1in} \alpha \neq 1
\end{align}
Suppose $s_1, s_2 \sim S(\alpha,1,1)$ i.i.d. For any  constants $c_1\geq 0, c_2\geq 0$, we have $c_1s_1 +c_2s_2 \sim S(\alpha,1,c_1^\alpha+c_2^\alpha)$. More generally, $\sum_{i=1}^N x_i s_i\sim S\left(\alpha,1,\sum_{i=1}^Nx_i^\alpha\right)$ if $s_i\sim S(\alpha,1,1)$ i.i.d.

%

There is a standard procedure to  sample from $S(\alpha,1,1)$~\cite{Article:Chambers_JASA76}. We  first generate an exponential random variable with mean 1, $w \sim \exp(1)$,  and a uniform random variable $u \sim unif \left(0, \pi\right)$, and then compute
\begin{align}\label{eqn_CMS}
\frac{\sin\left(\alpha u\right)}{\left[\sin u \cos\left(\alpha\pi/2 \right)
\right]^{\frac{1}{\alpha}}} \left[\frac{\sin\left( u - \alpha u\right)}{w}
\right]^{\frac{1-\alpha}{\alpha}} \sim S(\alpha,1,1)
\end{align}
In practice, we can replace the stable distribution with a heavy-tailed distribution in the domain of attractions~\cite{Book:Feller_II}, for example, $\frac{1}{\left[unif(0,1)\right]^{1/\alpha}}$. Again, we leave it as future work to use a sparsified design matrix.

\subsection{Data Streams and Linear Projection Methods}

The use of maximally-skewed stable random projections for nonnegative (dynamic) data stream computations was proposed in a line of work called {\em Compressed Counting (CC)}~\cite{Proc:Li_SODA09,Proc:Li_UAI09,Proc:Li_Zhang_COLT11}. Prior to CC, it was popular to use {\em symmetric stable random projections}~\cite{Article:Indyk_JACM06,Proc:Li_SODA08} in data stream computations.

In the standard {\em turnstile} data stream model~\cite{Article:Muthukrishnan_05}, at time $t$, an arriving stream element $(i_t, I_t)$ updates one entry of the data vector in a linear fashion: $x_{i_t}^{(t)} = x_{i_t}^{(t-1)}+I_t$. The dynamic nature of data streams makes computing the summary statistics, e.g., $\sum_{i=1}^N |x_i|^2$, and recovering the nonzero entries more challenging, especially if the streams arrive at high-speed (e.g., network traffic). Linear projections are naturally capable of  handling data streams. To see this,  suppose we denote the linear measurements as
\begin{align}
y_j^{(t)} = \sum_{i=1}^N x_i^{(t)} s_{ij},\ \ j = 1, 2, ..., M
\end{align}
When a new stream element $(i_t, I_t)$ arrives, we only need to update the measurement as
\begin{align}
y_j^{(t)} = y_j^{(t-1)} + I_t s_{i_t,j}, \ \  j = 1, 2, ..., M
\end{align}
The entries $s_{i_t,j}$ are re-generated as needed by using pseudo-random numbers~\cite{Proc:Nisan_STOC90}, i.e., no need to materialize the entire design matrix. This is the standard practice in data stream computations.

Here, we should mention  that this streaming model is actually very general. For example, the process of histogram-building can be viewed as a typical example of turnstile data streams. In machine learning, databases, computer vision, and NLP (natural language processing) applications, histogram-based features are  popular. In network applications, monitoring traffic histograms is an important mechanism for (e.g.,) anomaly detections~\cite{Proc:Feinstein_DARPA03}. Detecting (recovering) heavy components (e.g., so called ``elephant detection'') using compressed sensing is an active research topic in networks; see (e.g.,)~\cite{Proc:Zhao_IMC07,Proc:Lin_Globecom12,Proc:Wang_SIGCOMM12,Proc:Wang_Infocom12}.\\

For the rest of paper, we will drop the superscript $(t)$ in $y_j^{(t)}$ and $x_i^{(t)}$, while readers should keep in mind that our results are naturally applicable to data streams.

\subsection{The Proposed Algorithm and Main  Result}

For recovering a nonnegative signal $x_i\geq0$, $i = 1$ to $N$,  we  collect linear measurements $y_j = \sum_{i=1}^N x_i s_{ij}$, $j=1$ to $M$, where $s_{ij} \sim S(\alpha,1,1)$ i.i.d. In this paper, we focus on  $\alpha\in (0,0.5]$ and leave the study for $\alpha>0.5$ in future work.  At the decoding stage, we estimate the signal coordinate-wise:
\begin{align}
\hat{x}_{i,min} = \min_{1\leq j\leq M}\ y_j/s_{ij}
\end{align}
The number of measurements $M$  is chosen so that $\sum_{i=1}^N\mathbf{Pr}\left(\hat{x}_{i,min} -x_i \geq \epsilon \right)\leq \delta$ (e.g., $\delta=0.05$).\\

\noindent \textbf{Main Result}: \ \ When $\alpha\in(0,\ 0.5]$, it suffices to use  $M=(C_\alpha+o(1)) \epsilon^{-\alpha} \left(\sum_{i=1}^N x_i^\alpha\right)\log N/\delta$ measurements so that, with probability $1-\delta$, all coordinates will be recovered within $\epsilon$ additive precision, in one scan of the coordinates. The constant $C_\alpha=1$ when $\alpha\rightarrow0$ and $C_\alpha=\pi/2$ when $\alpha=0.5$. In particular, when $\alpha\rightarrow0$, the required number of measurements is essentially $M=K\log N/\delta$, where $K = \sum_{i=1}^N 1\{x_i\neq 0\}$ is the number of nonzero coordinates of the signal.\\

In the literature, it is  known that the sample complexity of compressed sensing using Gaussian design (i.e., $\alpha=2$) is essentially about $2K\log N/\delta$~\cite{Article:Donoho09,Proc:OMP_NIPS09}. This means our work already achieves smaller complexity with explicit constant, by requiring only one linear scan of the coordinates. Very encouragingly, it is perhaps not surprising that our method as presented in this paper is merely a tip of the iceberg and we expect a variety of followup works can be developed along this line. For example, it appears possible to further improve the algorithm by introducing iterations. It is also possible to sparsify the design matrix to  significantly speed up the processing (matrix-vector multiplication) and recovery.

\section{Preparation: Relevant Probability Results}

Our proposed algorithm utilizes  only the ratio statistics  $y_j/s_{ij}$ for recovery, while the observed data include more information, i.e.,  $(y_j, s_{ij})$ for $i = 1, 2, ..., N$, and $j = 1, 2, ..., M$. Thus, we first need to provide an explanation why we restrict ourselves to the ratio statistics. For convenience, we  define
\begin{align}
\theta = \left(\sum_{i=1}^N x_i^\alpha\right)^{1/\alpha},\ \ \
\theta_i = \left(\theta^\alpha- x_i^\alpha\right)^{1/\alpha}
\end{align}
and denote the probability density function of $s_{ij}\sim S(\alpha,1,1)$ by $f_S$. By a conditional probability argument, the joint density of $(y_j,s_{ij})$ can be shown to be $\frac{1}{\theta_i}f_S(s_{ij})f_S\left(\frac{y_j-x_is_{ij}}{\theta_i}\right)
\propto\frac{1}{\theta_i}f_S\left(\frac{y_j-x_is_{ij}}{\theta_i}\right)$. The MLE procedure amounts to finding $(x_i,\theta_i)$ to maximize the joint likelihood
\begin{align}\label{eqn_Lik}
L(x_i,\theta_i) =\prod_{j=1}^M\frac{1}{\theta_i}f_S\left(\frac{y_j-x_is_{ij}}{\theta_i}\right)
\end{align}
Interestingly, the following Lemma shows that  $L(x_i,\theta_i)$ approaches infinity at the poles $y_j - x_i s_{ij}=0$.
\begin{lemma}\label{lem_MLE}
The likelihood in (\ref{eqn_Lik})  approaches infinity, i.e., $L(x_i,\theta_i)\rightarrow +\infty$, if $y_j -x_i s_{ij} \rightarrow 0$, for any $j$, $1\leq j\leq M$.\\

\noindent\textbf{Proof}:\ \ See Appendix~\ref{app_lem_MLE}.$\hfill\square$ \\
\end{lemma}
The result in Lemma~\ref{lem_MLE}  suggests us to use only the ratio statistics $y_j/s_{ij}$ to recover $x_i$. By the property of stable distributions,
\begin{align}
\frac{y_j}{s_{ij}} = \frac{\sum_{t=1}^Nx_ts_{tj}}{s_{ij}}=x_i + \frac{\sum_{t\neq i}^Nx_ts_{tj}}{s_{ij}} = x_i + \theta_i \frac{S_2}{S_1}
\end{align}
where $\theta_i = \left(\sum_{t\neq i} x_i^\alpha\right)^{1/\alpha}$ and $S_1, S_2\sim S(\alpha,1,1)$ i.i.d. This motivates us to study the probability distribution of two independent stable random variables: $S_2/S_1$. For convenience, we define
\begin{align}
F_\alpha(t) = \mathbf{Pr}\left(\left({S_2}/{S_1}\right)^{\alpha/(1-\alpha)}\leq t\right), \ \ t\geq 0
\end{align}

\begin{lemma}\label{lem_F}

For any $t\geq 0$, $S_1, S_2 \sim S(\alpha,1,1)$, i.i.d.,
\begin{align}
&F_\alpha(t)
= \mathbf{Pr}\left(\left({S_2}/{S_1}\right)^{\alpha/(1-\alpha)}\leq t\right)= \frac{1}{\pi^2}\int_0^{\pi}\int_0^{\pi} \frac{1}{1+Q_\alpha/t}du_1 du_2
\end{align}
where
\begin{align}
Q_\alpha = \left[\frac{\sin\left(\alpha u_2\right)}{\sin\left(\alpha u_1\right)}\right]^{\alpha/(1-\alpha)}\left[\frac{\sin u_1}{\sin u_2
}\right]^{\frac{1}{1-\alpha}} \frac{\sin\left( u_2 - \alpha u_2\right)}{\sin\left( u_1 - \alpha u_1\right)}
\end{align}
In particular, closed-forms expressions are available when $\alpha\rightarrow0+$ or $\alpha = 0.5$:
\begin{align}
&\lim_{\alpha\rightarrow 0} F_\alpha(t) = \frac{1}{1+1/t},\hspace{0.5in}
F_{0.5}(t) = \frac{2}{\pi}\tan^{-1}\sqrt{t}
\end{align}
Moreover, for any  $t\in [0,\ 1]$, $0<\alpha_1\leq \alpha_2\leq 0.5$, we have
\begin{align}
&  \frac{1}{1+1/t}\leq  F_{\alpha_1}(t)  \leq F_{\alpha_2}(t)\leq \frac{2}{\pi}\tan^{-1}\sqrt{t}
\end{align}

\noindent\textbf{Proof:}\ See Appendix~\ref{app_lem_F}. Figure~\ref{fig_F} plots $F_\alpha(t)$ for selected $\alpha$ values.\hspace{0.1in} $\hfill\square$\\
\end{lemma}

\begin{figure}[h!]
\begin{center}
\includegraphics[width=3.5in]{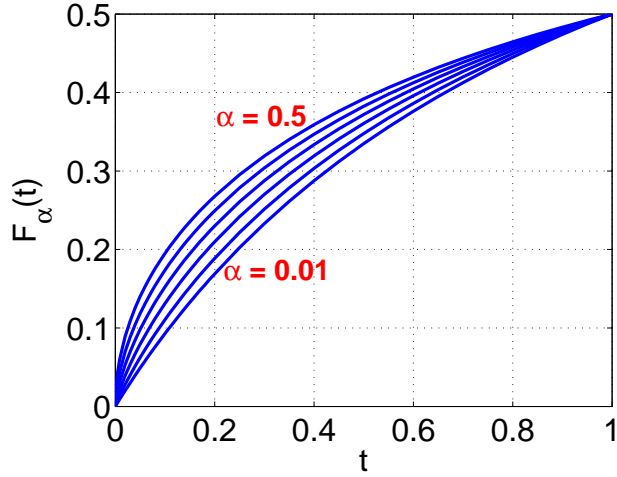}
\end{center}
\vspace{-0.25in}
\caption{$F_\alpha(t)$ for $t\in[0,\ 1 ]$ and $\alpha  = 0.01, 0.1, 0.2, 0.3, 0.4$ and $0.5$ (from bottom to top).}\label{fig_F}
\end{figure}

Lemma~\ref{lem_F} has proved that, when $\alpha\rightarrow0+$, $F_\alpha(t)$ is of order $t$, and when $\alpha=0.5$, $F_\alpha(t)$ is of order $\sqrt{t}$. Lemma~\ref{lem_F_order} provide a more general result that  $F_\alpha(t) =\Theta\left(t^{1-\alpha}\right)$.

\begin{figure}[h!]
\begin{center}
\includegraphics[width=3.5in]{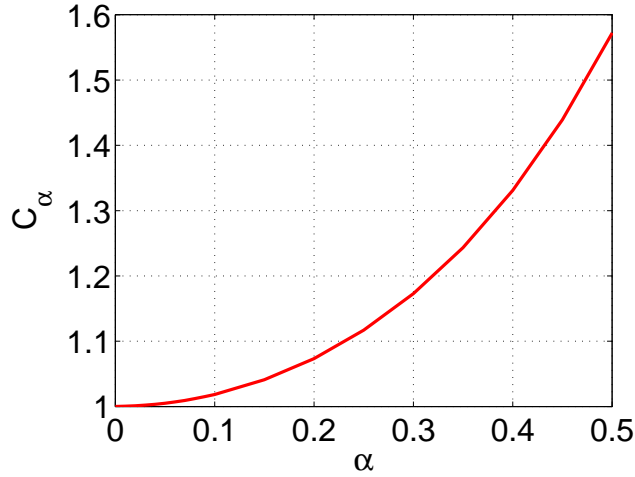}
\end{center}
\vspace{-0.25in}
\caption{The constant $C_\alpha$ as in Lemma~\ref{lem_F_order}. Numerically, it varies between 1 and $\pi/2$.}\label{fig_Calpha}
\end{figure}

\begin{lemma}\label{lem_F_order}
For $0\leq t<\alpha^{\alpha/(1-\alpha)}$ and $0<\alpha\leq 0.5$,
\begin{align}
 F_\alpha(t) =  \frac{t^{1-\alpha}}{C_\alpha + o(1)}
  \end{align}

\noindent\textbf{Proof:}\ See Appendix~\ref{app_lem_F_order}.\hspace{0.1in} $\hfill\square$
\end{lemma}
\noindent\textbf{Remarks for Lemma~\ref{lem_F_order}}:
\begin{itemize}
\item The result restricts $t<\alpha^{\alpha/(1-\alpha)}$. Here  $\alpha^{\alpha/(1-\alpha)}$ is monotonically decreasing in $\alpha$ and $0.5\leq \alpha^{\alpha/(1-\alpha)}\leq1$ for $\alpha\in(0, 0.5]$. Later we will show that our method only requires very small $t$.
\item The constant $C_\alpha$ can be numerically evaluated as shown in Figure~\ref{fig_Calpha}.
\item When $\alpha\rightarrow0+$, we have $F_{0+}(t) = \frac{1}{1+1/t} = t- t^2 + t^3  ... $. Hence $C_{0+} = 1$.
\item When $\alpha = 0.5$, we have $F_{0.5}(t) = \frac{2}{\pi}\tan^{-1}\sqrt{t} = \frac{2}{\pi}\left(t^{1/2} - t^{3/2}/3+...\right)$. Hence $C_{0.5} = \pi/2$.
    \end{itemize}

\newpage

To conclude this section, the next Lemma shows that the maximum likelihood estimator using the ratio statistics is actually the ``minimum estimator''.

\begin{lemma}\label{lem_ratio_MLE}
Use the ratio statistics, $y_j/s_{ij}$, $j=1$ to $M$. When  $\alpha\in(0,\ 0.5]$, the maximum likelihood estimator (MLE) of $x_i$ is the sample minimum
\begin{align}
\hat{x}_{i,min} = \min_{1\leq j\leq M} \ \frac{y_j}{s_{ij}}
\end{align}
\textbf{Proof:}\ See Appendix~\ref{app_lem_ratio_MLE}. $\hfill\square$\\
\end{lemma}

Lemma~\ref{lem_ratio_MLE} largely explains our proposed algorithm. In the next section, we analyze the error probability of $\hat{x}_{i,min}$ and its sample complexity bound.

\section{Error Probability,  Sample Complexity  Bound, and Bias Analysis}

The following Lemma concerns the tail probability of the estimator $\hat{x}_{i,min}$. Because $\hat{x}_{i,min}$ always over-estimates $x_i$, we only need to provide a one-sided error probability bound.
\begin{lemma}\label{lem_err}

\begin{align}
\mathbf{Pr}\left(\hat{x}_{i,min} -x_i \geq \epsilon \right) =& \left[1-F_\alpha\left(\left(\epsilon/\theta_i\right)^{\alpha/(1-\alpha)}\right)\right]^M\\\label{eqn_Err_bound}
\leq& \left[\frac{1}{1+\left(\epsilon/\theta_i\right)^{\alpha/(1-\alpha)}}  \right]^M
\end{align}
For $0<\alpha\leq 0.5$ and  $\epsilon/\theta_i<\alpha$,
\begin{align}\label{eqn_Err_order}
\mathbf{Pr}\left(\hat{x}_{i,min} -x_i \geq \epsilon \right) =\left[1-\Theta\left(\epsilon^\alpha/\theta_i^{\alpha}\right)\right]^M
\end{align}
In particular, when $\alpha = 0.5$,
\begin{align}
\mathbf{Pr}\left(\hat{x}_{i,min} -x_i \geq \epsilon, \alpha = 0.5 \right) = \left[1-\frac{2}{\pi}\tan^{-1}\sqrt{\frac{\epsilon}{\theta_i}}\right]^M
\end{align}
\textbf{Proof:}\ \ Recall $\frac{y_j}{s_{ij}} = x_i + \theta_i\frac{S_{2}}{S_1}$ and $\hat{x}_{i,min} = \min_{1\leq j\leq M}\ \ \frac{y_j}{s_{ij}}$. We have
\begin{align}\notag
&\mathbf{Pr}\left(\hat{x}_{i,min}> x_i + \epsilon\right)=\mathbf{Pr}\left(\frac{y_j}{s_{ij}} > x_i + \epsilon,\ 1\leq j \leq M\right)\\\notag
=&\left[ \mathbf{Pr}\left(\frac{S_2}{S_1} > \frac{\epsilon}{\theta_i}\right)\right]^M= \left[1-F_\alpha\left(\left(\epsilon/\theta_i\right)^{\alpha/(1-\alpha)}\right)\right]^M
\end{align}
The rest of the proof follows from Lemma~\ref{lem_F} and Lemma~\ref{lem_F_order}.$\hfill\square$
\end{lemma}
\textbf{Remark for Lemma~\ref{lem_err}:}\ The probability bound (\ref{eqn_Err_bound}) is  convenient to use. However, it is conservative in that it does not give the right order unless $\alpha$ is small (i.e., when $\alpha/(1-\alpha)\approx \alpha$). In comparison, (\ref{eqn_Err_order}) provides the exact order, which will be useful for analyzing the precise sample complexity of our proposed algorithm. As shown in Lemma~\ref{lem_F_order}, $F_\alpha(t)= \Theta(t^{1-\alpha})$ holds for relatively small $t<\alpha^{\alpha/(1-\alpha)}$. In our case, $t = \left(\epsilon/\theta_i\right)^{\alpha/(1-\alpha)}$, i.e., the result requires $\epsilon/\theta_i <\alpha$, or $\epsilon^\alpha/\theta_i^\alpha = \epsilon^\alpha/(\sum_{l\neq i}^N x_l^\alpha) <\alpha^\alpha$. When $\alpha\rightarrow0$, this means we need $1/K<1$, which is virtually always true. For larger $\alpha$, the relation $\epsilon^\alpha/(\sum_{l\neq i}^N x_l^\alpha) <\alpha^\alpha$ should hold for any reasonable settings.\\

\begin{theorem}\label{thm_complexity}
To ensure $\sum_{i=1}^N\mathbf{Pr}\left(\hat{x}_{i,min} -x_i \geq \epsilon \right)\leq \delta$, it suffices to choose $M$ by
\begin{align}\label{eqn_MF}
M\geq \frac{\log N/\delta}{-\log \left[1-F_\alpha\left(\left(\epsilon/\theta\right)^{\alpha/(1-\alpha)}\right)\right]}
\end{align}
where $F_\alpha$ is defined in Lemma~\ref{lem_F}. If $\epsilon/\theta<1$, then it  suffices to use
\begin{align}\label{eqn_M0}
M\geq \frac{\log N/\delta}{\log \left[1+\left(\epsilon/\theta\right)^{\alpha/(1-\alpha)}  \right]}
\end{align}
which is sharp when $\alpha\rightarrow0$.  In general, for $\alpha\in(0,\ 0.5]$ and $\epsilon/\theta<\alpha$, the (sharp) bound can be written as,
\begin{align}\label{eqn_MC}
M\geq \left(C_\alpha+o(1)\right)\left(\frac{\theta}{\epsilon}\right)^{\alpha}\log N/\delta
\end{align}
where the constant $C_\alpha$ is the same constant in Lemma~\ref{lem_F_order}.

When $\alpha=0.5$ and $\epsilon/\theta<1$, a precise bound exists:
\begin{align}\label{eqn_M05}
M\geq \frac{\pi}{2}\sqrt{\frac{\theta}{\epsilon}} \log N/\delta
\end{align}

\noindent\textbf{Proof:}\ \ The result (\ref{eqn_MF}) follows from    Lemma~\ref{lem_err}, (\ref{eqn_M0}) from  Lemma~\ref{lem_F}, (\ref{eqn_MC}) from Lemma~\ref{lem_F_order}.

We provide more details for the proof of the more precise bound (\ref{eqn_M05}).  When $\alpha=0.5$,
\begin{align}\notag
M\geq\frac{\log N/\delta}{-\log \left[1-\frac{2}{\pi}\tan^{-1}\sqrt{\frac{\epsilon}{\theta}}\right]}
\end{align}
which can be simplified to be $M\geq \frac{\pi}{2}\sqrt{\frac{\theta}{\epsilon}} \log N/\delta$, using the fact that $-\log\left(1-\frac{2}{\pi}\tan^{-1}(z)\right) \geq \frac{2}{\pi}z, \forall z\in[0,\ 1]$.  To see this inequality,  we can check
\begin{align}\notag
\frac{\partial}{\partial z}\left(-\log(1-\frac{2}{\pi}\tan^{-1}(z)) - \frac{2}{\pi}z\right) = \frac{\frac{2}{\pi}}{\left(1-\frac{2}{\pi}\tan^{-1}{z}\right)(1+z^2)} -\frac{2}{\pi}
\end{align}
It suffices to show
\begin{align}\notag
 z^2 - \frac{2}{\pi}\tan^{-1}z -\frac{2}{\pi}z^2\tan^{-1}z \leq 0
\end{align}
which is true because the equality holds when $z=0$ or $z=1$, and
\begin{align}\notag
\frac{\partial^2}{\partial z^2}\left( z^2 - \frac{2}{\pi}\tan^{-1}z -\frac{2}{\pi}z^2\tan^{-1}z\right) = 2-\frac{2}{\pi}\left(2z\tan^{-1}z +\frac{2z}{1+z^2}\right)>0
\end{align}
This completes the proof.$\hfill\square$
\end{theorem}
\textbf{Remarks for Theorem~\ref{thm_complexity}:}\ \ The convenient bound (\ref{eqn_M0}) is only sharp for $\alpha\rightarrow0$.  For example, when $\alpha=0.5$, $\alpha/(1-\alpha) =1$, but the true order should be in terms of $\sqrt{\epsilon}$ instead of $\epsilon$. The  other bound (\ref{eqn_MC}) provides the precise order, where the constant $C_\alpha$ is the same as in Lemma~\ref{lem_F_order}.  The fact that the complexity is proportional to $\epsilon^{-\alpha}$ is  important and presents a substantial improvement over the previous $O\left(\epsilon^{-1}\right)$ result in Count-Min sketch~\cite{Article:Cormode_05}. For example, if we let $\alpha\rightarrow0$, then  $\left(\frac{\theta}{\epsilon}\right)^{\alpha}\rightarrow K$. In other words, the complexity for exact $K$-sparse recovery is essentially $K\log N/\delta$ and the constant is basically 1. We will comment more on the choice of $\alpha$ later in the paper. \\

To conclude this section, we provide the analysis for the \text{bias}.  The minimum estimator $\hat{x}_{i,min}$ is biased and  it always over-estimates $x_i$. The following Lemma evaluates the bias precisely.
\begin{lemma}\label{lem_bias}
\begin{align}
&E\left(\hat{x}_{i,min}\right) =x_i + \theta_iD_{M,\alpha}\\
&D_{M,\alpha} = \int_{0}^\infty \left[1-F_\alpha\left(t^{\alpha/(1-\alpha)}\right)\right]^Mdt
\end{align}
In particular, when $\alpha=0.5$,
\begin{align}
&D_{M,\alpha=0.5} = M(M-1)\frac{4}{\pi^2}\sum_{j=0}^\infty
 \frac{(-1)^j\pi^{2j}}{(M+2j-2)(2j)!} B_{2j} -1
\end{align}
where $B(a,b) = \int_0^1 t^{a-1}(1-t)^{b-1}dt$ is the Beta function and $B_j$ is the Bernoulli number satisfying
\begin{align}\notag
\frac{t}{e^t-1} = \sum_{j=0}^\infty B_j \frac{t^j}{j!}=\sum_{j=0}^\infty B_{2j} \frac{t^{2j}}{(2j)!} - \frac{t}{2}
\end{align}
e.g., $B_0=1$, $B_1=-1/2$, $B_2 = 1/6$, $B_4=-1/30$, $B_6=1/42$, $B_8=-1/30$, $B_{10} = 5/66$, ...\\

\noindent\textbf{Proof:}\ See Appendix~\ref{app_lem_bias}. Figure~\ref{fig_DM05} plots the $D_{M,\alpha}$ for $\alpha=0.5$. $\hfill\square$\\

\end{lemma}

\begin{figure}[h!]
\begin{center}
\includegraphics[width=3.5in]{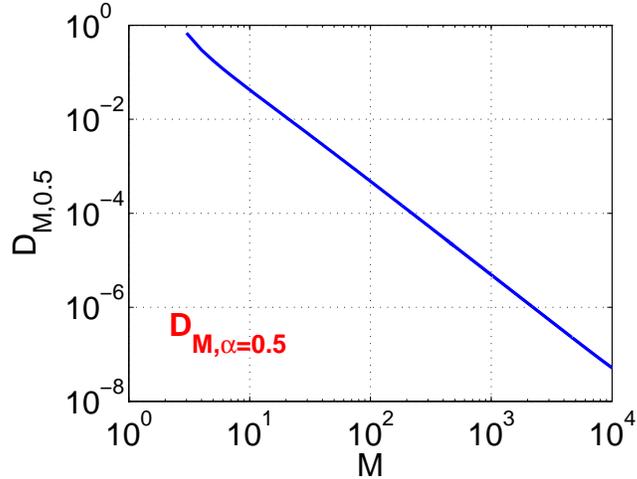}
\end{center}
\vspace{-0.25in}
\caption{The constant $D_{M,\alpha=0.5}$ for the bias analysis in Lemma~\ref{lem_bias}.}\label{fig_DM05}
\end{figure}

\newpage

\section{Experiments}\label{sec_exp}

Our proposed algorithm for sparse recovery is  simple and  requires  merely one scan of the coordinates. Our theoretical analysis provides the sharp sample complexity bound with the constant (i.e., $C_\alpha$)  specified (e.g., Figure~\ref{fig_Calpha}).   It is nevertheless  still interesting to include an experimental study. All experiments presented in this study were conducted in Matlab on a workstation with 256GB memory. We did not make special effort to optimize our code for efficiency.

We  compare our proposed method with two popular L1 decoding packages: {\em L1Magic}~\cite{Report:L1Magic} and {\em SPGL1}~\cite{Article:SPGL1}\footnote{We must specify some parameters in order to achieve sufficient accuracies. For { L1Magic}, we use the following Matlab script:\begin{verbatim} l1eq_pd(x0, Afun, Atfun, y,1e-3,100,1e-8,1000);\end{verbatim} For { SPGL1}, after consulting the author of~\cite{Article:SPGL1}, we used the following  script: \begin{verbatim} opts = spgSetParms('verbosity',0); opts.optTol=1e-6;opts.decTol=1e-6;spg_bp(A, y, opts);\end{verbatim} However, it looks for $N=10,000,000$ and $K=10$ we probably should reduce the tolerance further (which would increase the computational time substantially). Here, we would like to thank  the authors of both~\cite{Report:L1Magic} and ~\cite{Article:SPGL1} for discussions on this issue.}, on simulated data. Although it is certainly not our intension to compare the two L1 decoding solvers, we decide to present the results of both. While it is known that SPGL1 can often be  faster than L1Magic, we observe that in some cases SPGL1 could not achieve the desired accuracy. On the other hand, SPGL1 better uses  memory and can handle larger problems than L1Magic.  \\

In each simulation, we randomly select $K$ out $N$ coordinates and set their values ($x_i$) to be 1. The other $N-K$ coordinates are set to be 0. To simulate the design matrix $\mathbf{S}$, we generate  two random matrices: $\{u_{ij}\}$ and $\{w_{ij}\}$, $i = 1$ to $N$, $j = 1$ to $M$, where $u_{ij}\sim unif(0,\pi)$ and $w_{ij}\sim exp(1)$, i.i.d. Then we apply the formula (\ref{eqn_CMS}) to generate $s_{ij} \sim (\alpha,1,1)$, for $\alpha = 0.04$ to 0.5, spaced at 0.01. We also use the same $u_{ij}$ and $w_{ij}$ to generate standard Gaussian ($N(0,1)$) variables for the design matrix used by { L1Magic} and { SPGL1}, based on  the  interesting fact:\ $-\sqrt{2}\cos(u_{ij})\sqrt{w_{ij}}\sim N(0,1)$.\\


In this experimental setting, since $K = \sum_{i=1}^N x_i^\alpha$, the sample complexity of our algorithm is essentially $M = C_\alpha K/\epsilon^\alpha \log N/\delta$, where $C_{0+}=1$ and $C_{0.5}=\pi/2 \approx 1.6$. In our simulations, we choose $M$ by two options: (i) $M = K\log N/\delta$; (ii) $M= 1.6K\log N/\delta$, where $\delta = 0.01$.\\

We compare our method with { L1Magic} and { SPGL1} in terms of the decoding times and the recovery errors. The (normalized) recovery error is defined as
\begin{align}
error = \sqrt{\frac{\sum_{i=1}^N (x_i-\text{estimated } x_i)^2}{\sum_{i=1}^N x_i^2}}
\end{align}

\subsection{$M = K\log N/\delta$}

\begin{figure}[h!]
\begin{center}
\includegraphics[width=3in]{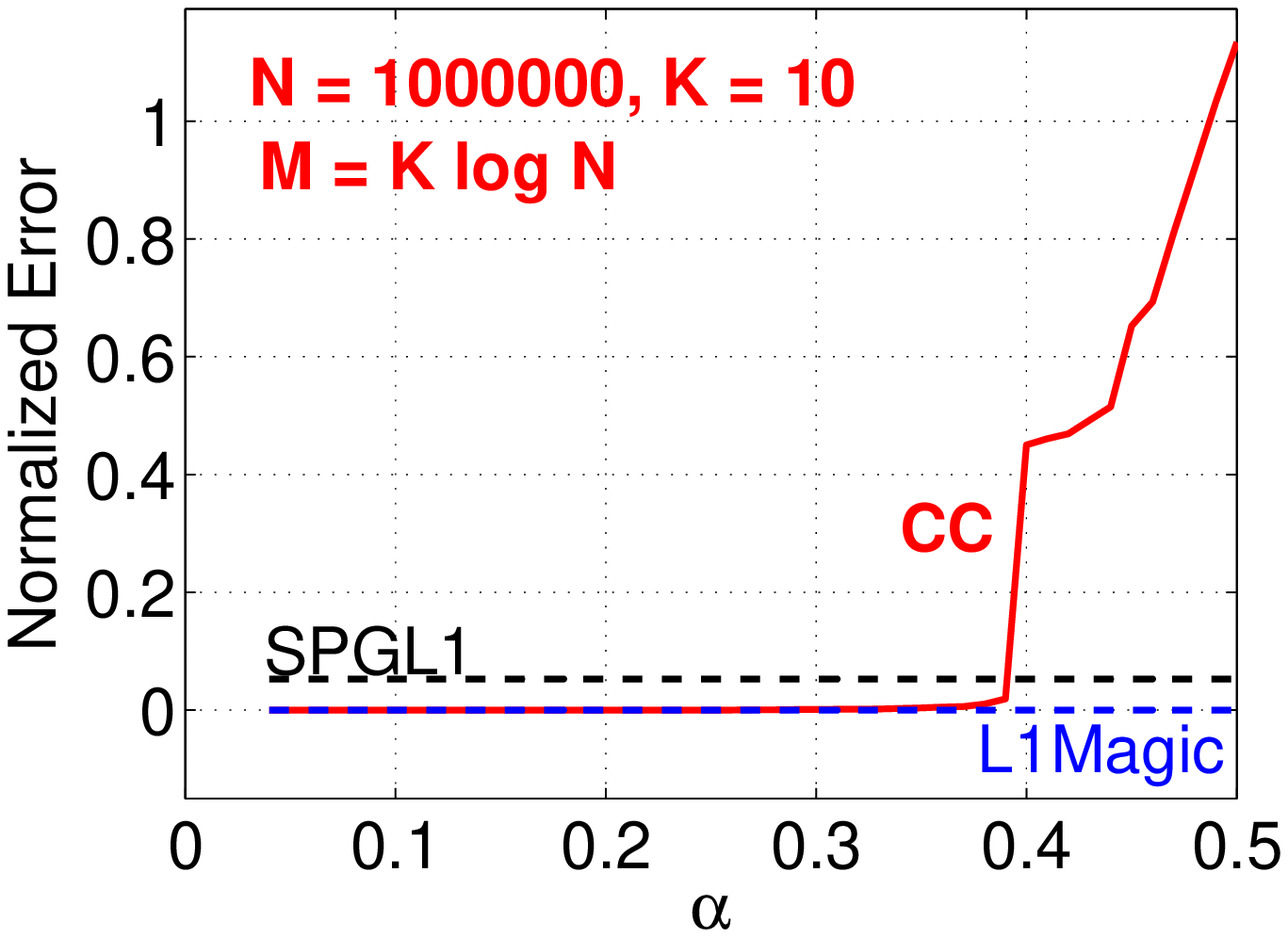}\hspace{0.2in}
\includegraphics[width=3in]{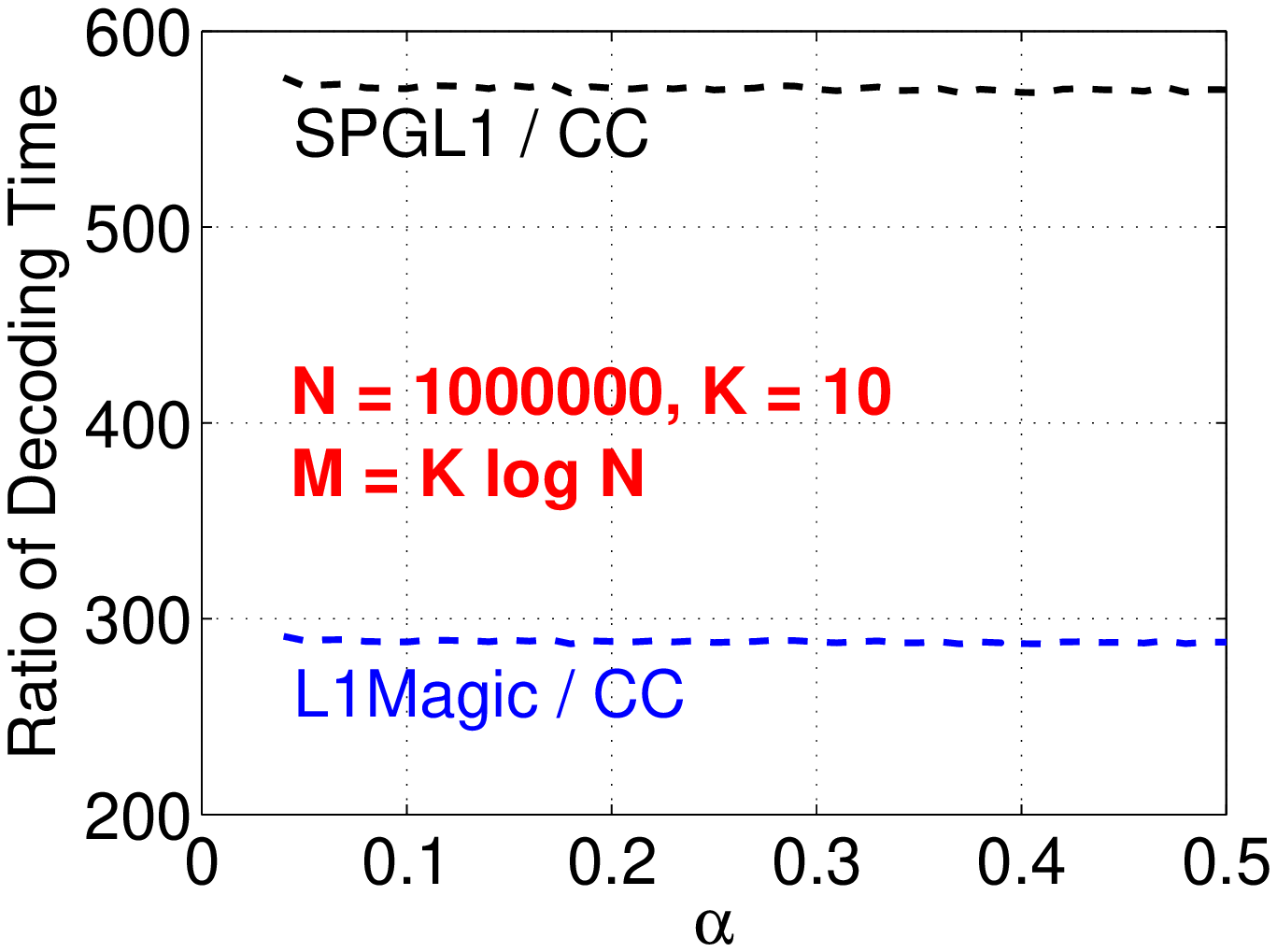}
\end{center}
\vspace{-0.25in}
\caption{Experiments for comparing our proposed algorithm  (labeled ``CC'')  with SPGL1~\cite{Article:SPGL1} and L1Magic~\cite{Report:L1Magic}, for $N=1,000,000$, $K=10$, and $M=K\log N/\delta$ (where $\delta=0.01$). For each $\alpha$ (from 0.04 to 0.5 spaced at 0.01), we conduct simulations 100 times and report the median results. In the \textbf{left panel}, our proposed method  (solid curve) produces very accurate recovery results for $\alpha<0.38$. For larger $\alpha$ values, however, the errors become large. This is expected because when $\alpha=0.5$, the required number of samples should be $(\pi/2)K\log N/\delta$ instead of $K\log N/\delta$.  In this case, L1Magic also produces accuracy recovery results. Note that for all methods, we report the top-$K$ entries of the recovered signal as the estimated nonzero entries. In the \textbf{right panel}, we plot the ratios of the decoding times. Basically, SPGL1 package uses about 580 times more time than our proposed method (which requires only one  scan), and L1Magic package needs about 290 times more time than our method.  }\label{fig_L1N1000000K10B1}\vspace{0.2in}

\end{figure}

Figure~\ref{fig_L1N1000000K10B1} presents the recovery errors (left panel) and ratios of the decoding times (right panel), for $N=1,000,000$, $K=10$, and $M=K\log N/\delta$ (where $\delta=0.01$). The results confirm that our proposed method is computationally very efficient and is capable of producing accurate recovery for $\alpha<0.38$. \\

\begin{figure}[h!]
\begin{center}
\includegraphics[width=3in]{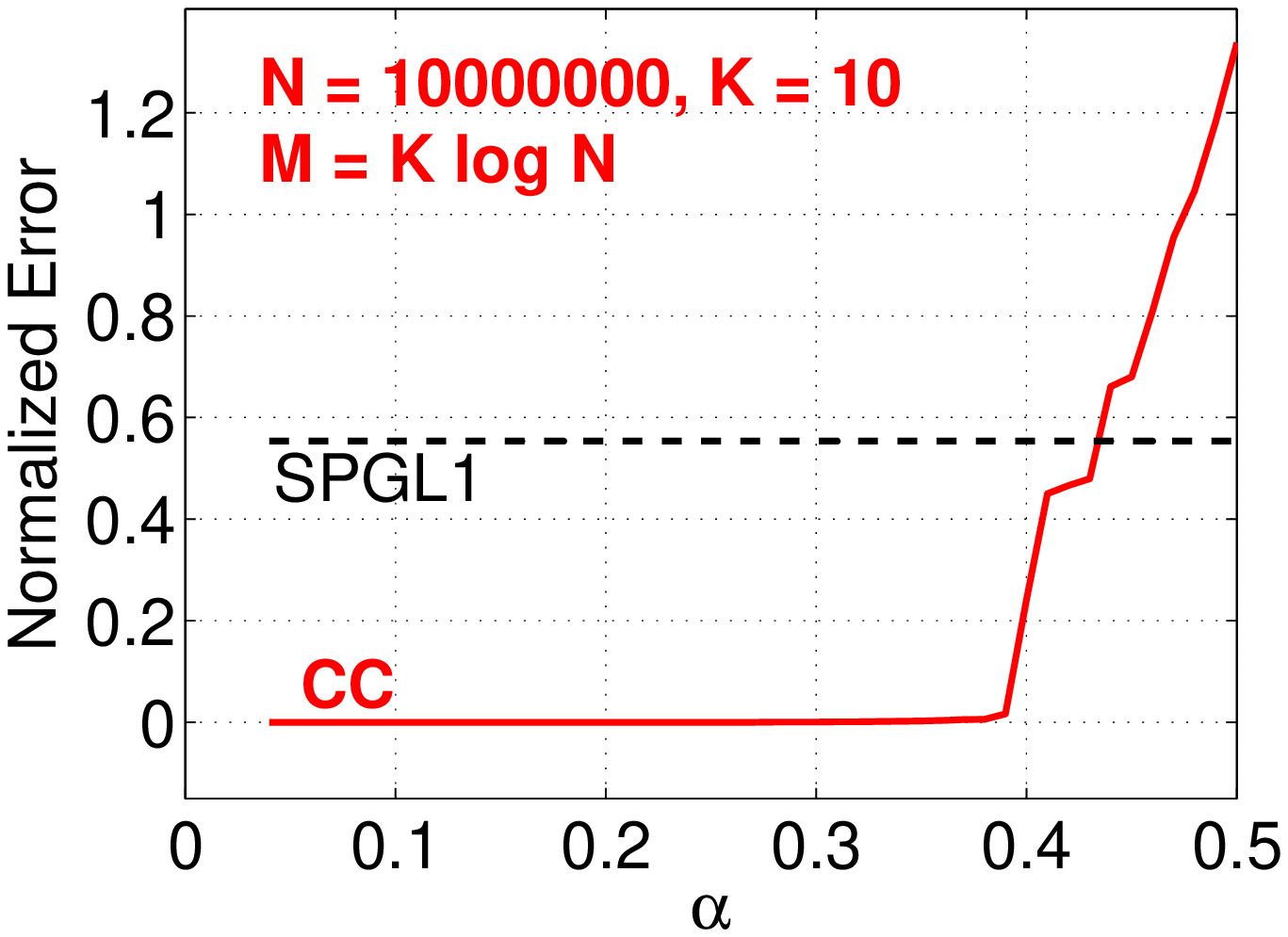}\hspace{0.2in}
\includegraphics[width=3in]{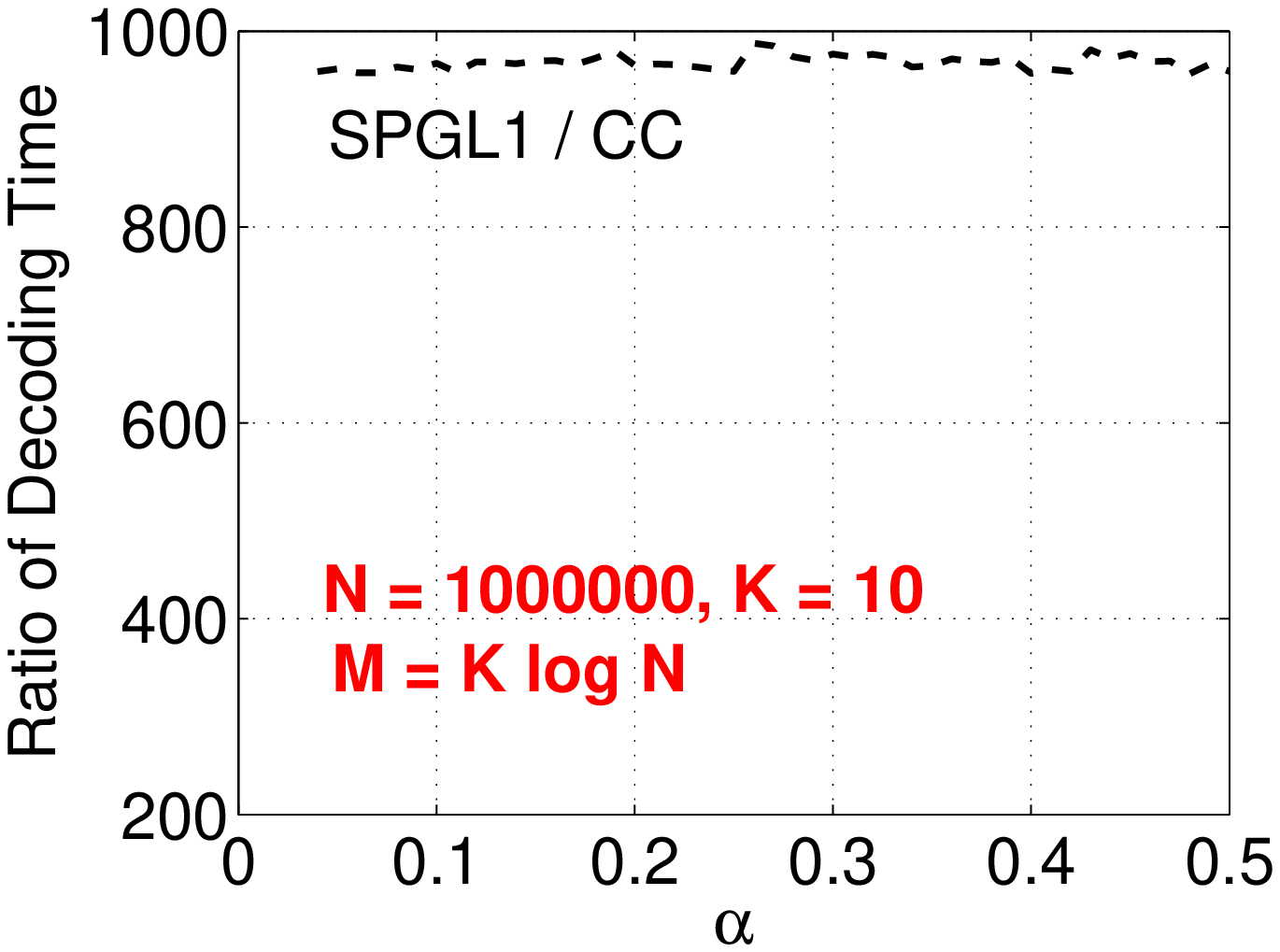}
\end{center}
\vspace{-0.25in}
\caption{Experiments for comparing our proposed algorithm  (CC) with SPGL1 and L1Magic, for $N=10,000,000$, $K=10$, and $M=K\log N/\delta$ (where $\delta=0.01$). See the caption of Figure~\ref{fig_L1N1000000K10B1} for more details. In this larger problem, we can not run L1Magic as the program simply halts without making progress.}\label{fig_SPGL1N10000000K10B1}
\end{figure}

Figure~\ref{fig_SPGL1N10000000K10B1} presents the results for a larger problem, with $N=10,000,000$ and $K=10$. Because we can not run L1Magic in this case, we only present the comparisons with SPGL1. Again, our method is computationally very efficient and produces accurate recovery for about $\alpha<0.38$.

\newpage

For $\alpha$ close to 0.5, we need to increase the number of measurements, as shown in the theoretical analysis.

\subsection{$M = 1.6K\log N/\delta$}

\begin{figure}[h!]
\begin{center}
\includegraphics[width=3in]{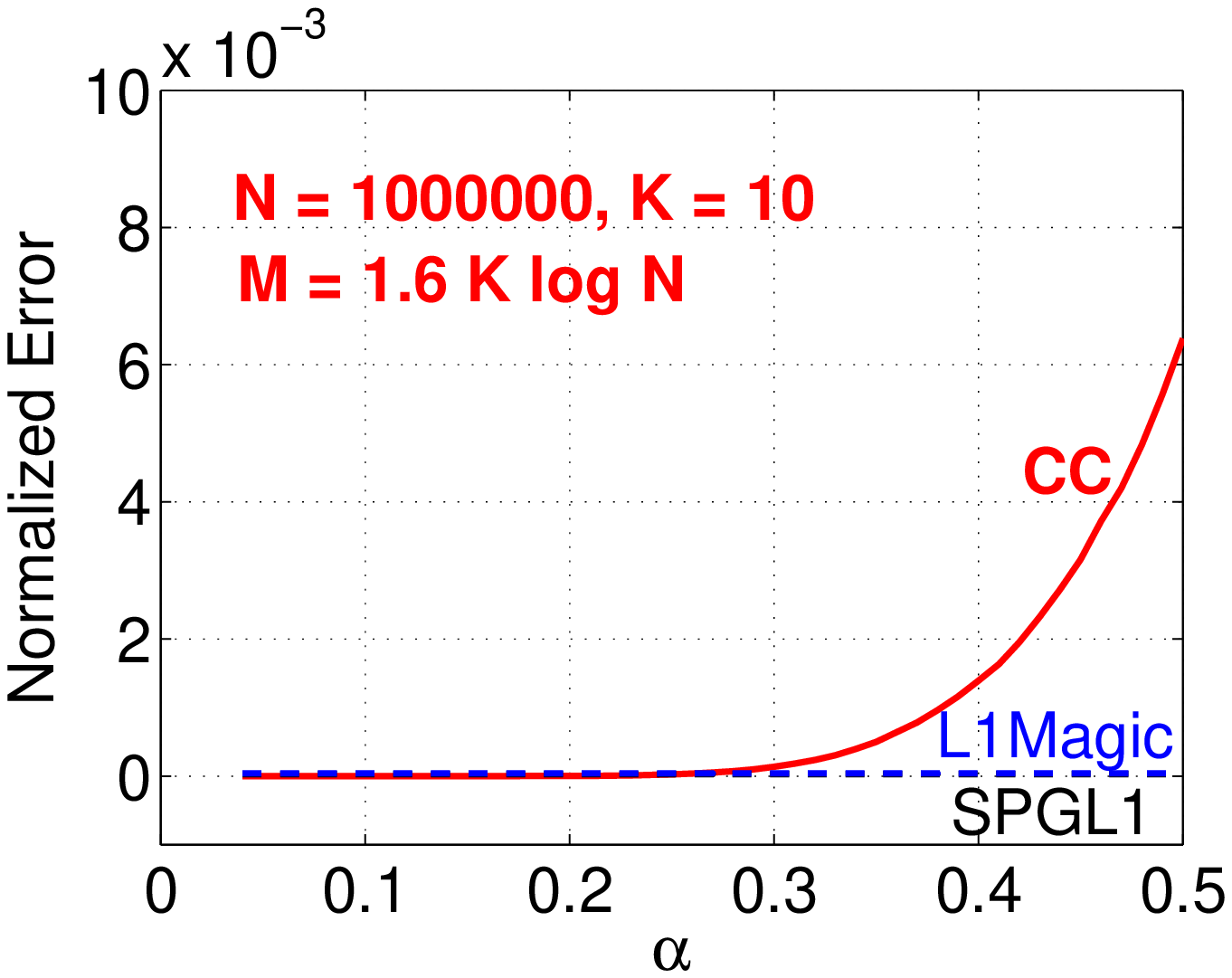}\hspace{0.2in}
\includegraphics[width=3in]{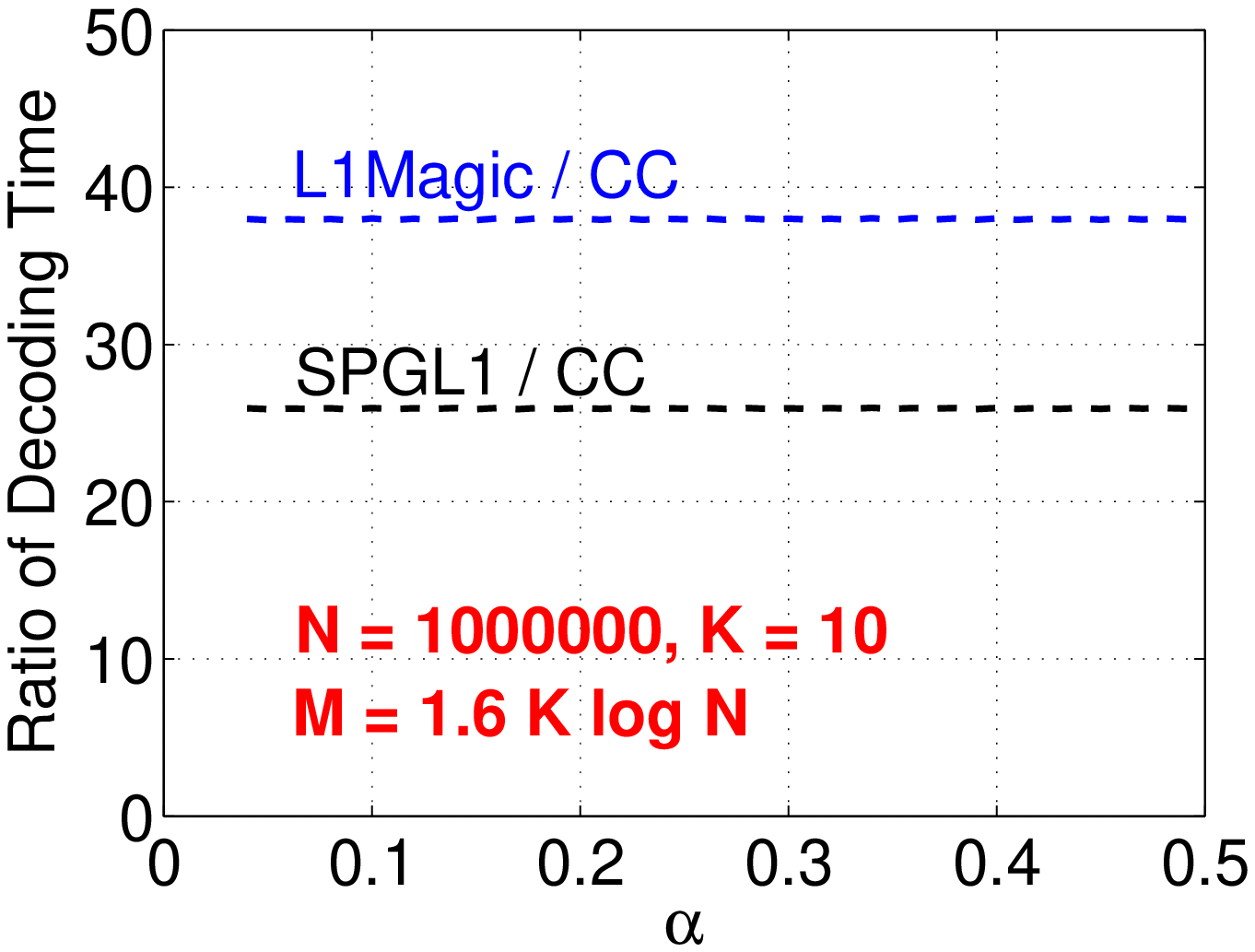}
\end{center}
\vspace{-0.25in}
\caption{Experiments for comparing our proposed algorithm  (CC) with SPGL1 and L1Magic, for $N=1,000,000$, $K=10$, and $M=1.6K\log N/\delta$ (where $\delta=0.01$). Again, for each $\alpha$, we conduct simulations 100 times and report the median results. In the \textbf{left panel}, our proposed method  (solid curve) produces accurate recovery results, although the errors increase with increasing $\alpha$ (the maximum error is around 0.007). In the \textbf{right panel}, we can see that in this case, our method is only, respectively, 27 times and 39 times faster than SPGL1 and L1Magic. We should mention that we did not make special effort to optimize our Matlab code for efficiency. }\label{fig_L1N1000000K10B1.6}
\end{figure}

\begin{figure}[h!]
\begin{center}
\includegraphics[width=3in]{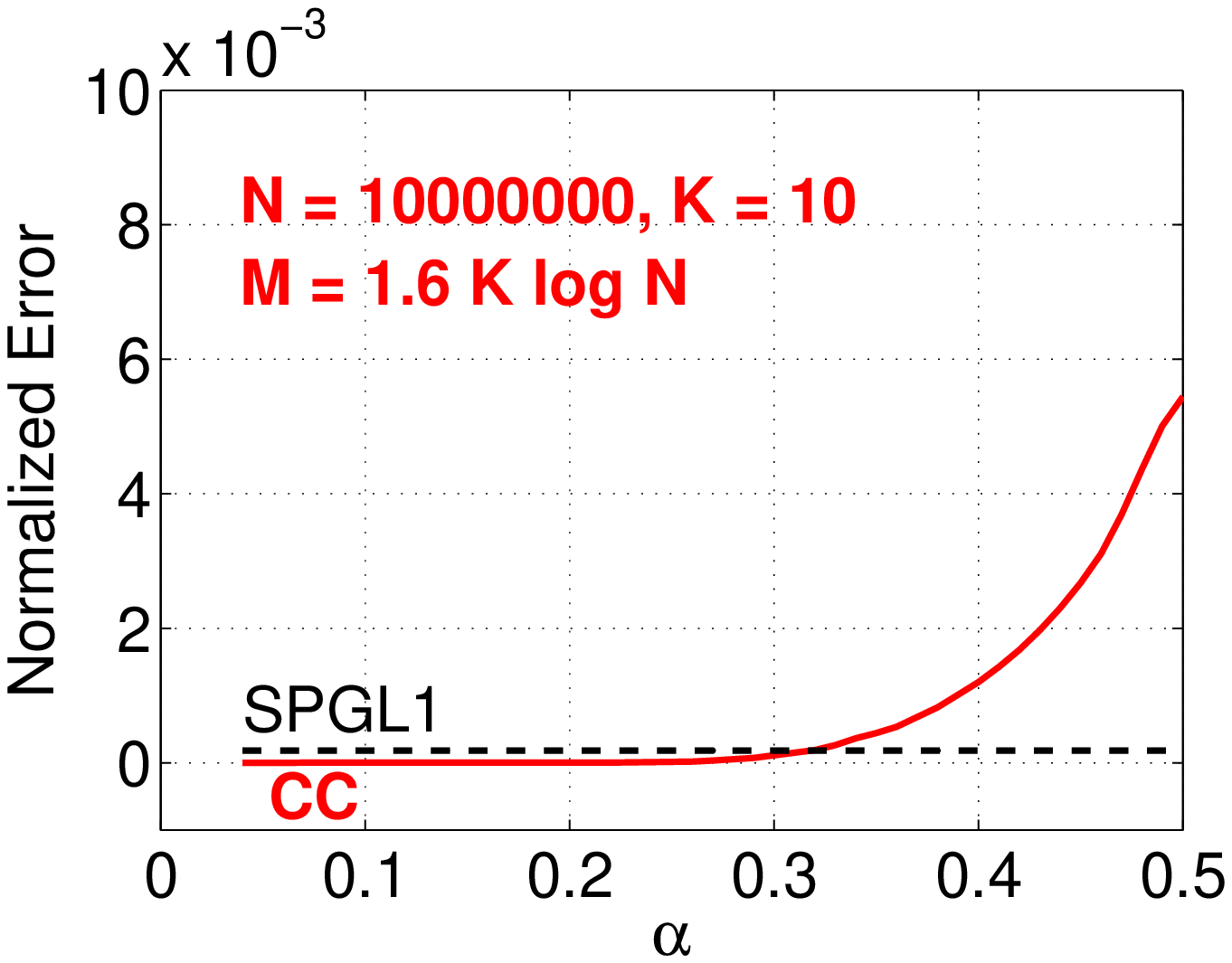}\hspace{0.2in}
\includegraphics[width=3in]{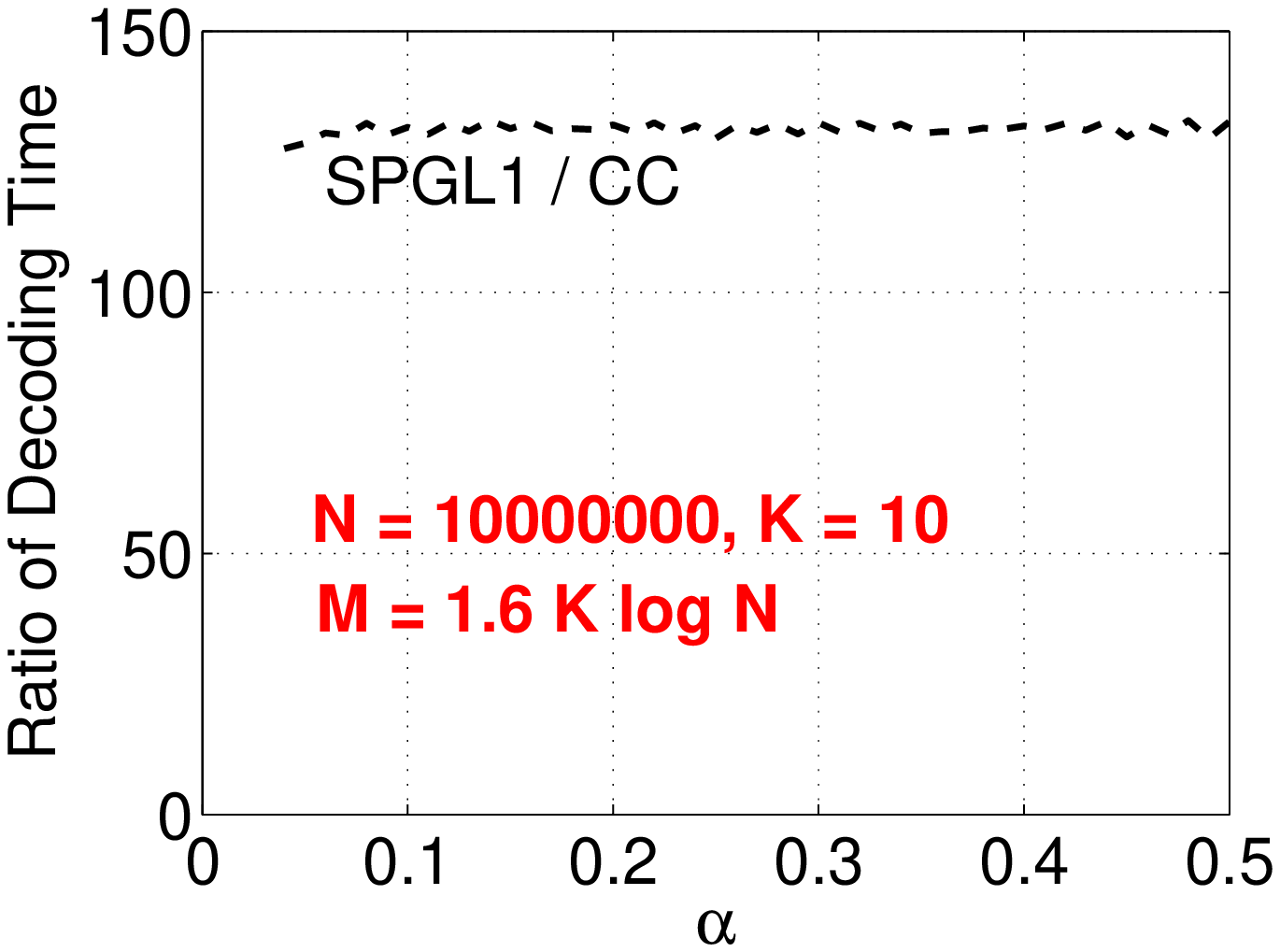}
\end{center}
\vspace{-0.25in}
\caption{Experiments for comparing our proposed algorithm  (CC) with SPGL1 and L1Magic, for $N=10,000,000$, $K=10$, and $M=K\log N/\delta$ (where $\delta=0.01$).}\label{fig_SPGL1N10000000K10B1.6}
\end{figure}

To study the behavior as $\alpha$ approaches 0.5, we increase the number of measurements to $M=1.6K\log N/\delta$. Figure~\ref{fig_L1N1000000K10B1.6} and Figure~\ref{fig_SPGL1N10000000K10B1.6} present the experimental results for $N=1,000,000$ and $N=10,000,000$, respectively. Interestingly, when $\alpha=0.5$, our algorithm still produces accurate recovery results (with the normalized errors around 0.007), although the results at smaller $\alpha$ values are even more accurate. In the next subsection (Section~\ref{subsec_bias}), we will experimentally show that the recovery accuracy can be further improved by a bias-correction procedure as derived Lemma~\ref{lem_bias}.

\newpage

\subsection{Bias-Correction}\label{subsec_bias}

As analyzed in Lemma~\ref{lem_bias}, the minimum estimator $\hat{x}_{i,min}$ is slightly biased: $E\left(\hat{x}_{i,min}\right) = x_i + \theta_i D_{M,\alpha}$, where the constant $D_{M,\alpha}$ can be pre-computed and tabulated for each $M$ and $\alpha$ (e.g., Figure~\ref{fig_DM05} for $D_{M,\alpha}$ with $\alpha=0.5$). We also need to estimate $\theta_i$, for which we resort the estimator in the prior work on Compressed Counting~\cite{Proc:Li_UAI09}. For example, for $\alpha=0.5$, the bias-corrected estimator is
\begin{align}\label{eqn_minc}
\alpha=0.5:\hspace{0.2in} \hat{x}_{i,min,c}  = \hat{x}_{i,min} - \left[\left(1-\frac{3}{4M}\right)\sqrt{\frac{M}{\sum_{j=1}^M1/y_j}} - \sqrt{\hat{x}_{i,min}}\right]^2D_{M,0.5}
\end{align}

\begin{figure}[h!]
\begin{center}
\includegraphics[width=3.5in]{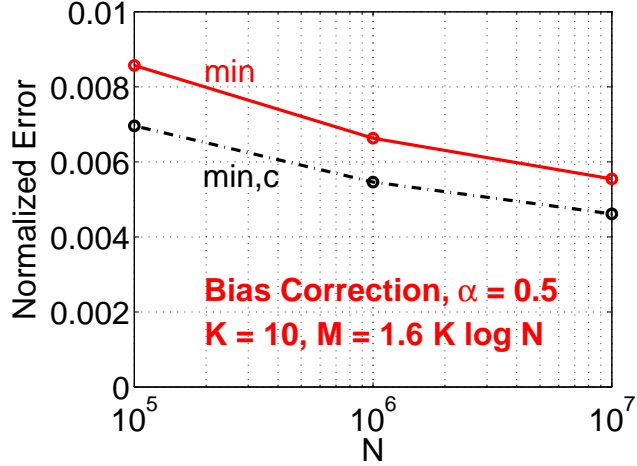}
\end{center}
\vspace{-0.25in}
\caption{Bias correction for further improving the minimum estimator of our proposed algorithm at $\alpha=0.5$. In this experiment, we choose $K=10$, $M=1.6K\log N/\delta$, and $N=10^5, 10^6, 10^7$. In each simulation, we use the original minimum estimator $\hat{x}_{i,min}$ together with the bias-corrected estimator $\hat{x}_{i,min,c}$ as in (\ref{eqn_minc}). We can see that the bias-correction step does improve the accuracy, as the dashed error curve ($\hat{x}_{i,min,c}$) is lower than the solid error curve ($\hat{x}_{i,min}$).  }\label{fig_CCBiasK10S1B1.6}\vspace{0.2in}
\end{figure}

As verified in Figure~\ref{fig_CCBiasK10S1B1.6}, the bias-corrected estimator (\ref{eqn_minc}) does improve the original minimum estimator.

\subsection{Robustness against Measurement Noise}

\begin{figure}[h!]
\begin{center}

\mbox{\includegraphics[width=2.7in]{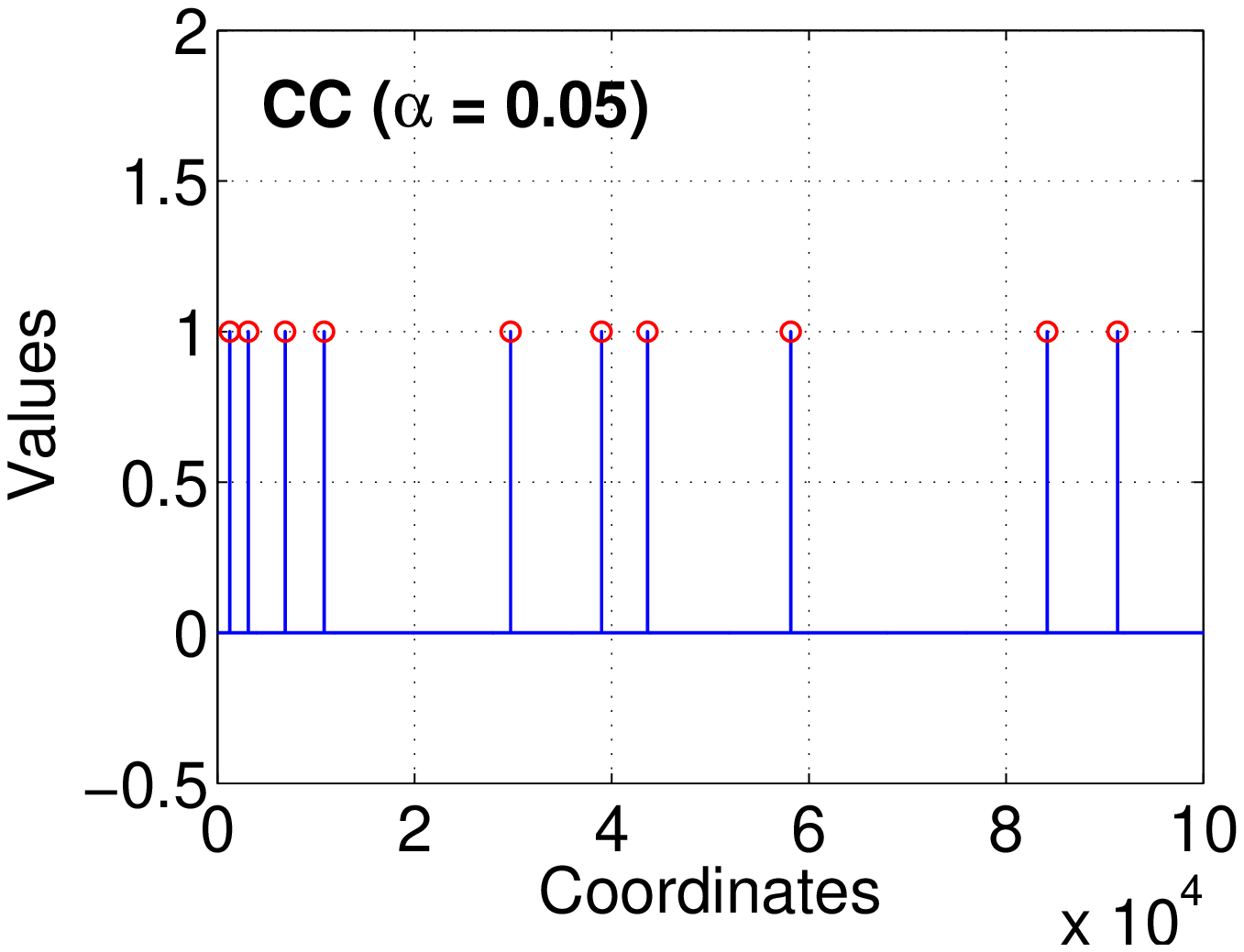}\hspace{0.2in}
\includegraphics[width=2.7in]{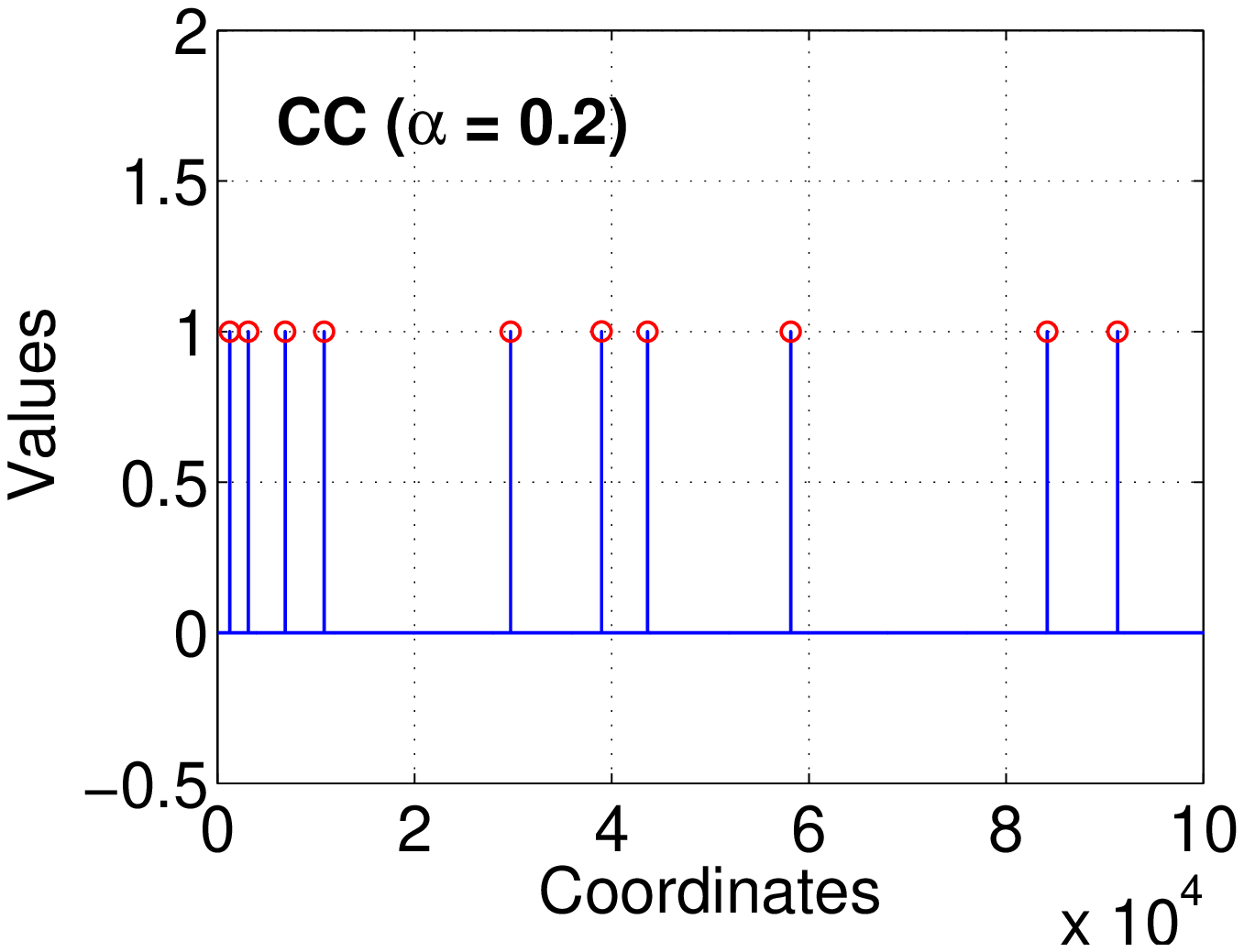}
}

\mbox{\includegraphics[width=2.7in]{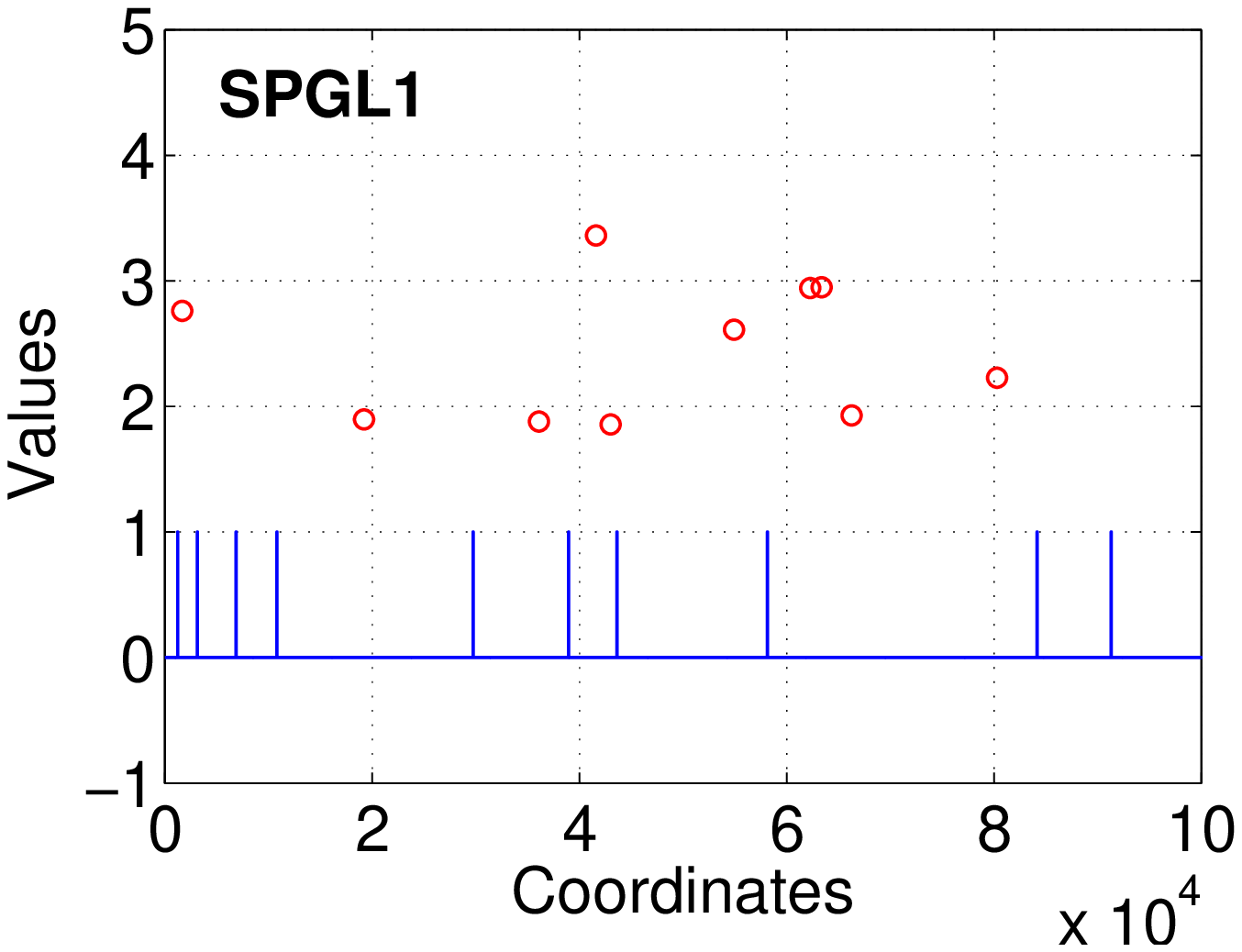}\hspace{0.2in}
\includegraphics[width=2.7in]{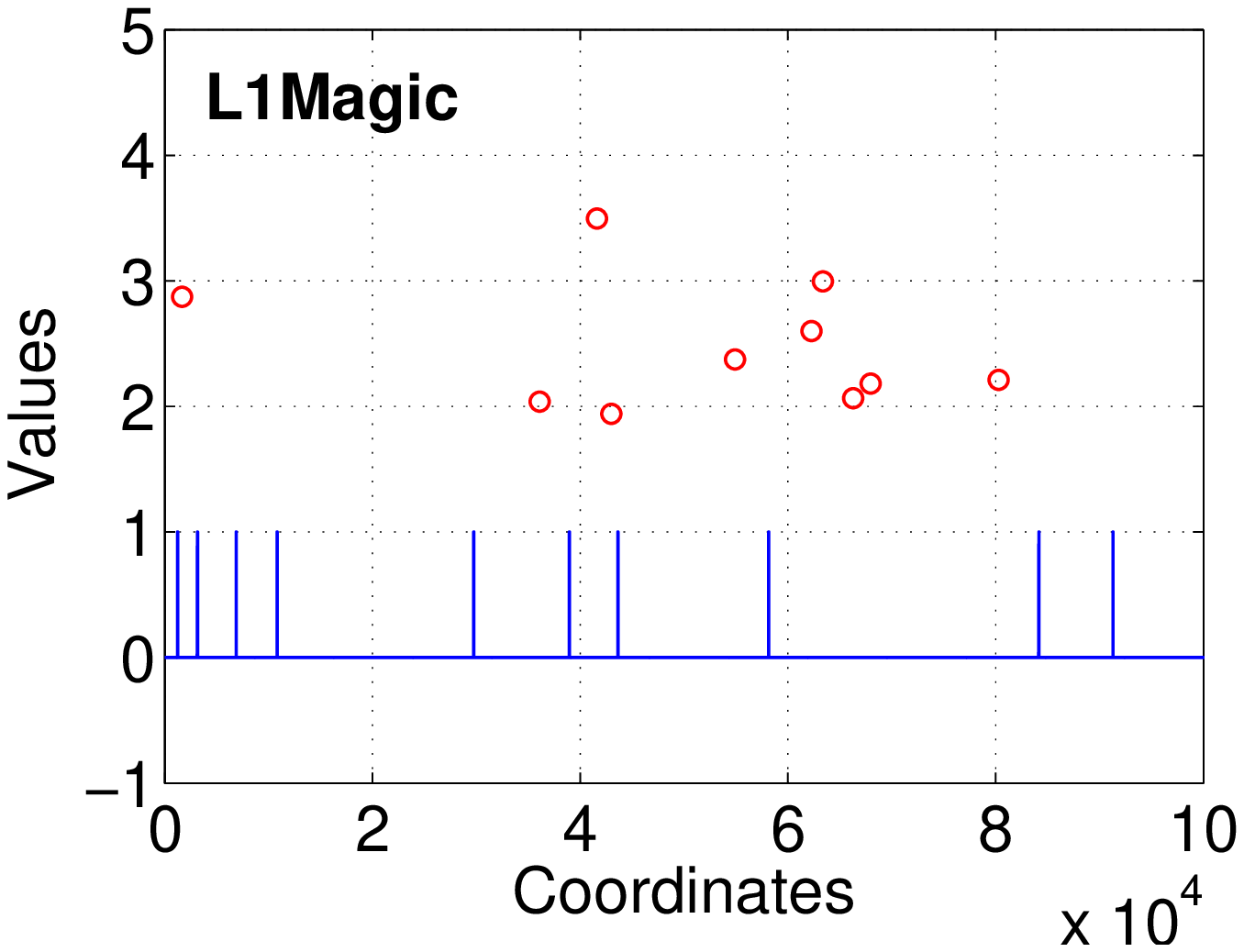}
}

\end{center}
\vspace{-0.25in}
\caption{In this experiment, we choose $N=100,000$, $K=10$, $M=K\log N/\delta$ (with $\delta=0.01$). We add noises to the measurements: $y_j = \sum_{i=1}^Nx_i s_{ij} + n_j$, where $n_j \sim N(0,\sigma^2)$ i.i.d. In this example, we let $\sigma^2 = N\sigma_0^2$ where $\sigma_0=0.1$. We then run our proposed algorithm (CC, for $\alpha=0.05$ and $\alpha=0.2$) and L1 solvers (L1Magic and SPGL1). In each panel, the solid  straight lines stand for the values of the nonzero entries and the (red) circles are the recovered nonzero coordinates reported by algorithms.   Clearly, our proposed algorithm is essentially indifferent to the measurement noises while the two L1 solvers are not robust against measurement noises.}\label{fig_noise}
\end{figure}

Figure~\ref{fig_noise} presents an experimental study to illustrate that our proposed algorithm is robust against usual measurement noise model:
\begin{align}
y_j  = \sum_{i=1}^N x_i s_{ij} + n_j,\hspace{0.2in}\text{where }\  n_j \sim N\left(0, \sigma^2\right), \ j=1, 2, ..., M, i.i.d.
\end{align}
where the  noise $n_j$ can, for example, come from transmission channel after collecting the measurements.

It is actually very intuitive to understand why our proposed algorithm can be robust against measurement noises. Using the ratio statistics, we have  $\frac{y_j}{s_{ij}} = x_i + \theta_i \frac{S_2}{S_1} + \frac{n_j}{S_1}$.  Because $S_1$ is very heavy-tailed, the noise in terms of $n_j/S_1$,  has essentially no impact. In this paper, we only provide the intuitive explanation and leave a formal analysis in future work.

\newpage

\section{Discussion and Future Work}

While our proposed algorithm for sparse recovery based on compressed counting (maximally-skewed $\alpha$-stable random projections) is simple and fast, it is clear that the work presented in this paper is merely a tip of the iceberg. We expect many interesting related research problems will arise. \\

One important issue is the choice of $\alpha$. In this paper, our analysis focuses on $\alpha\in(0,\ 0.5]$ and our theoretical results show that smaller $\alpha$ values lead to better performance. The natural question is: why can we simply use a very small $\alpha$? There are numerical issues which prevents us from using a too small $\alpha$.

For convenience, consider the approximate mechanism for generating $S(\alpha,1,1)$ by using $1/U^{1/\alpha}$, where $U\sim unif(0,1)$ (based on the theory of domain of attractions and generalized central limit theorem). If $\alpha=0.04$, then we have to compute $(1/U)^{25}$, which may potentially create numerical problems. In our Matlab simulations, we use $\alpha \in [0.04,\ 0.5]$ and we do not notice obvious numerical problems even with $\alpha=0.04$ as shown in Section~\ref{sec_exp}. However, if a device (e.g., camera or other hand-held device)  has more limited precision and memory, then we expect that we must use a larger $\alpha$. Fortunately, our experiments in  Section~\ref{sec_exp} show that the performance is not too sensitive to $\alpha$. For example, in our experimental setting, the recovery accuracies are very good for $\alpha<0.38$ even when we choose the sample size $M$ based on $\alpha\rightarrow0$.\\

Among many potential future research problems, we list a few examples as follows:
\begin{itemize}
\item When the signal can have both positive and negative components, we need to use symmetric stable random projections.
\item The sample complexity of our algorithm is $O\left(\epsilon^{-\alpha}\right)$. For small $\alpha$, the value of $\epsilon^{-\alpha}$ is close to 1 even for small $\epsilon$, for example $0.01^{-0.04} = 1.2$.  If a device allows the use of very small $\alpha$, then we expect some iteration scheme might be able to substantially reduce the required number of measurements.
\item In this paper, we focus on dense design matrix. In KDD'07, the work on ``very sparse random projections''~\cite{Proc:Li_KDD07} showed that one can significantly sparsify the design matrix without hurting the performance in estimating summary statistics. We expect that it is also possible to use sparsified design matrix in our framework for sparse recovery. However, since recovering summary statistics is in general an easier task than recovering all the coordinates, we expect there will be nontrivial analysis for (e.g.,) deciding the level of sparsity without hurting the recovery results.

\item Another interesting issue is the coding of the measurements $y_j$, which is a practical issue because storing and transmitting the measurements can be costly. Recently, there is work~\cite{Report:RPCode2013} for coding Gaussian random projections in the context of search and learning. We expect some ideas in~\cite{Report:RPCode2013} might be borrowed for sparse recovery.
\end{itemize}

\section{Conclusion}

We develop a new compressed sensing algorithm using {\em Compressed Counting (CC)} which is based on {\em maximally-skewed $\alpha$-stable random projections}. Our method produces accurate recovery of nonnegative sparse signals and our procedure is  computationally very efficient. The cost  is just  one linear scan of the coordinates. Our theoretical analysis provides the sharp complexity bound. While our preliminary results are encouraging, we expect many promising future research problems can be pursued in this line of work.

\newpage\clearpage
{

}

\newpage\clearpage

\appendix

\section{Proof of Lemma~\ref{lem_MLE}}\label{app_lem_MLE}

For $S\sim S(\alpha,1,1)$, the sampling approach in (\ref{eqn_CMS}) provides a method to compute its CDF
\begin{align}\notag
F_S(s) =& \mathbf{Pr}\left(\frac{\sin\left(\alpha u\right)}{\left[\sin u \cos\left(\alpha\pi/2 \right)
\right]^{\frac{1}{\alpha}}} \left[\frac{\sin\left( u - \alpha u\right)}{w}
\right]^{\frac{1-\alpha}{\alpha}}\leq s\right)\\\notag
=&\mathbf{Pr}\left(\frac{\left[\sin\left(\alpha u\right)\right]^{\alpha/(1-\alpha)}}{\left[\sin u \cos\left(\alpha\pi/2 \right)
\right]^{\frac{1}{1-\alpha}}} \left[\frac{\sin\left( u - \alpha u\right)}{w}
\right]\leq s^{\alpha/(1-\alpha)}\right)\\\notag
=&\frac{1}{\pi}\int_0^\pi \exp\left\{-\frac{\left[\sin\left(\alpha u\right)\right]^{\alpha/(1-\alpha)}}{\left[\sin u \cos\left(\alpha\pi/2 \right)
\right]^{\frac{1}{1-\alpha}}} \left[\frac{\sin\left( u - \alpha u\right)}{s^{\alpha/(1-\alpha)}}
\right] \right\}du\\\notag
=&\frac{1}{\pi}\int_0^\pi \exp\left\{-q_\alpha(u) s^{-\alpha/(1-\alpha)}\right\}du
\end{align}
and the PDF
\begin{align}\notag
f_S(s)
=&\frac{1}{\pi}\int_0^\pi \exp\left\{-q_\alpha(u) s^{-\alpha/(1-\alpha)}\right\}q_\alpha(u)\alpha/(1-\alpha)s^{-\alpha/(1-\alpha)-1} du
\end{align}
Hence,
\begin{align}\notag
&\frac{1}{\theta_i}f_S\left(\frac{y_j-x_is_{ij}}{\theta_i}\right) \\\notag
=&\frac{\alpha/(1-\alpha)}{\pi}\int_0^\pi q_\alpha(u)\exp\left\{-q_\alpha(u) \left(\frac{\theta_i}{y_j-x_is_{ij}}\right)^{\alpha/(1-\alpha)}\right\} \left(\frac{\theta_i}{y_j-x_is_{ij}}\right)^{\alpha/(1-\alpha)} \frac{1}{\left(y_j-x_is_{ij}\right)}du
\end{align}
Therefore, the likelihood $L(x_i,\theta_i)\rightarrow+\infty$ if $y_j - x_is_{ij}\rightarrow 0$, provided $\theta_i/(y_j-x_is_{ij})\rightarrow const$. Note that here we can choose $\theta_i$ and $x_i$ to maximize the likelihood.

\section{Proof of Lemma~\ref{lem_F}}\label{app_lem_F}

Since $S_1, S_2\sim S(\alpha,1,1)$, i.i.d., we know that
\begin{align}\notag
&S_1 = \frac{\sin\left(\alpha u_1\right)}{\left[\sin u_1 \cos\left(\alpha\pi/2 \right)
\right]^{\frac{1}{\alpha}}} \left[\frac{\sin\left( u_1 - \alpha u_1\right)}{w_1}
\right]^{\frac{1-\alpha}{\alpha}},\\\notag
&S_2 = \frac{\sin\left(\alpha u_2\right)}{\left[\sin u_2 \cos\left(\alpha\pi/2 \right)
\right]^{\frac{1}{\alpha}}} \left[\frac{\sin\left( u_2 - \alpha u\right)}{w_2}
\right]^{\frac{1-\alpha}{\alpha}}
\end{align}
where $u_1, u_2 \sim  uniform \left(0, \pi\right)$, $w_1, w_2 \sim \exp(1)$, $u_1, u_2, w_1, w_2$ are independent. Thus, we can write
\begin{align}\notag
&\left({S_2}/{S_1}\right)^{\alpha/(1-\alpha)} = Q_\alpha\frac{w_1}{w_2},\\\notag
&Q_\alpha = \left[\frac{\sin\left(\alpha u_2\right)}{\sin\left(\alpha u_1\right)}\right]^{\alpha/(1-\alpha)}\left[\frac{\sin u_1}{\sin u_2
}\right]^{\frac{1}{1-\alpha}} \frac{\sin\left( u_2 - \alpha u_2\right)}{\sin\left( u_1 - \alpha u_1\right)}
\end{align}

Using properties of exponential distributions, for any $t\geq 0$,
\begin{align}\notag
F_\alpha(t) = \mathbf{Pr}\left(\left({S_2}/{S_1}\right)^{\alpha/(1-\alpha)}\leq t\right) = \mathbf{Pr}\left(Q_\alpha w_1/w_2\leq t\right) =
E\left(\frac{1}{1+Q_\alpha/t}\right) = \frac{1}{\pi^2}\int_0^{\pi}\int_0^{\pi} \frac{1}{1+Q_\alpha/t}du_1 du_2
\end{align}

When $\alpha\rightarrow0+$, $Q_\alpha\rightarrow 1$ point-wise. By dominated convergence, $F_{0+}(t) = \frac{1}{1+1/t}$.

When $\alpha = 0.5$, $Q_\alpha$ can be simplified to be
\begin{align}\notag
Q_{0.5} = \left[\frac{\sin\left(u_2/2\right)}{\sin\left( u_1/2\right)}\right]\left[\frac{\sin u_1}{\sin u_2
}\right]^{2} \frac{\sin\left( u_2/2 \right)}{\sin\left( u_1/2\right)} = \frac{\cos^2\left(u_1/2\right)}{\cos^2\left( u_2/2\right)}
\end{align}
which can be used to obtain the closed-form expression for $F_{0.5}(t)$:
\begin{align}\notag
F_{0.5}(t)  =& \frac{1}{\pi^2}\int_0^{\pi}\int_0^{\pi} \frac{1}{1+Q_{0.5}/t}du_1 du_2\\\notag
=&\frac{1}{\pi^2}\int_0^{\pi}\int_0^{\pi} \frac{1}{1+ \frac{\cos^2\left(u_1/2\right)}{t\cos^2\left( u_2/2\right)}}du_1 du_2\\\notag
=&\frac{4}{\pi^2}\int_0^{\pi/2}\int_0^{\pi/2} \frac{1}{1+ b\cos^2\left(u_1\right)}du_1 du_2,\hspace{0.2in} b = \frac{1}{t\cos^2\left( u_2\right)}\\\notag
=&\frac{4}{\pi^2}\int_0^{\pi/2} \left.\frac{-1}{\sqrt{1+b}}\tan^{-1}\left(\sqrt{1+b} \frac{\cos u_1}{\sin u_1}\right)\right|_{0}^{\pi/2} du_2\\\notag
=&\frac{2}{\pi}\int_0^{\pi/2} \frac{1}{\sqrt{1+\frac{1}{t}\sec^2u_2}} du_2\\\notag
=&\frac{2}{\pi}\int_0^{1} \frac{1}{\sqrt{1+\frac{1}{t}- z^2 }} dz
=\frac{2}{\pi}\int_0^{1/\sqrt{1+1/t}} \frac{1}{\sqrt{1- z^2 }} dz\\\notag
=&\frac{2}{\pi}\sin^{-1}\left(1/\sqrt{1+1/t}\right)
= \frac{2}{\pi}\tan^{-1}\sqrt{t}
\end{align}
%
%
%
%
%
%

To show $F_\alpha(t) \geq 1/(1+1/t)$ for any $t\in[0,\ 1]$, we first note that the equality holds when $t=0$ and $t=1$. To see the latter case, we write $Q_\alpha = q_2/q_1$, where $q_1$ and $q_2$ are i.i.d. When $t=1$, $F_\alpha(t) = E\left(1/(1+q_2/q_1)\right) = E\left(\frac{q_1}{q_1+q_2}\right) = \frac{1}{2}$ by symmetry.

It remains to show $F_\alpha(t)$ is monotonically increasing in $\alpha$ for fixed $t\in[0,\ 1]$. For convenience, we define $q_\alpha(u)$ and $g_\alpha(u)$, where
\begin{align}\notag
&Q_\alpha = q_\alpha(u_2)/q_\alpha(u_1), \hspace{0.2in} q_\alpha(u) = \left[\sin\left(\alpha u\right)\right]^{\alpha/(1-\alpha)}\left[\sin u\right]^{\frac{-1}{1-\alpha}} \sin\left( u - \alpha u\right)
\end{align}
\begin{align}\notag
g_\alpha(u) = \frac{\partial \log q_\alpha(u)}{\partial \alpha } = \frac{\cos\alpha u}{\sin \alpha u} \frac{\alpha u}{1-\alpha} + \frac{1}{(1-\alpha)^2}\log\sin\alpha u - \frac{1}{(1-\alpha)^2}\log\sin u - u\frac{\cos(u-\alpha u)}{\sin(u-\alpha u)}
\end{align}
We can  check that both $q_\alpha(u)$ and $g_\alpha(u)$ are monotonically increasing in $u\in[0, \pi]$.
%
%

\begin{align}\notag
&\frac{\partial g_\alpha(u)}{\partial u} =\frac{-\alpha}{\sin^2 \alpha u} \frac{\alpha u}{1-\alpha} +\frac{\cos\alpha u}{\sin \alpha u} \frac{\alpha }{1-\alpha} + \frac{\alpha}{(1-\alpha)^2}\frac{\cos\alpha u}{\sin\alpha u} - \frac{1}{(1-\alpha)^2}\frac{\cos u}{\sin u} - \frac{\cos(u-\alpha u)}{\sin(u-\alpha u)}+ \frac{(1-\alpha)u}{\sin^2(u-\alpha u)}\\\notag
=&\left\{\frac{(1-\alpha)u}{\sin^2(u-\alpha u)}-\frac{\alpha}{\sin^2 \alpha u} \frac{\alpha u}{1-\alpha} \right\} +\left\{\frac{\cos\alpha u}{\sin \alpha u} \frac{\alpha }{1-\alpha}- \frac{\cos(u-\alpha u)}{\sin(u-\alpha u)}\right\} + \left\{\frac{\alpha}{(1-\alpha)^2}\frac{\cos\alpha u}{\sin\alpha u} - \frac{1}{(1-\alpha)^2}\frac{\cos u}{\sin u}\right\}
\end{align}
We consider three terms (in curly brackets) separately and show they are all $\geq 0$ when $\alpha\in[0,\ 0.5]$.

For the first term,
\begin{align}\notag
\frac{(1-\alpha)u}{\sin^2(u-\alpha u)}-\frac{\alpha}{\sin^2 \alpha u} \frac{\alpha u}{1-\alpha} \geq 0\Longleftrightarrow \frac{1-\alpha}{\sin((1-\alpha)u)} \geq \frac{\alpha}{\sin \alpha u}
\Longleftrightarrow (1-\alpha)\sin\alpha u  -\alpha\sin((1-\alpha)u)\geq 0
\end{align}
where the last inequality holds because the derivative (w.r.t. $u$) is $(1-\alpha)\alpha\cos\alpha u  - (1-\alpha)\alpha\cos((1-\alpha)u)\geq 0$. For the second term, it suffices to show
\begin{align}\notag
&\frac{\partial}{\partial u} \left\{\alpha\cos\alpha u \sin(u-\alpha u)-(1-\alpha)\sin\alpha u\cos(u-\alpha u)\right\}\geq 0\\\notag
\Longleftrightarrow& -\alpha^2\sin\alpha u \sin(u-\alpha u)+(1-\alpha)^2\sin\alpha u\sin(u-\alpha u)\geq 0
\end{align}
For the third term, it suffices to show
\begin{align}\notag
\alpha\sin u\cos\alpha u-\cos u \sin\alpha u\geq 0 \Longleftrightarrow \alpha\sin(u-\alpha u) + (1-\alpha)\cos u \sin \alpha u\geq 0
\end{align}
Thus, we have proved the monotonicity of $g_\alpha(u)$ in $u\in[0,\ \pi]$, when $\alpha\in[0,\ 0.5]$.

To prove the monotonicity of $q_\alpha(u)$ in $u$, it suffices to check if its logarithm is monotonic, i.e.
\begin{align}\notag
\frac{\partial}{\partial u} \log q_a\alpha(u) = \frac{1}{1-\alpha}\left(\alpha^2\frac{\cos\alpha u}{\sin\alpha u} + (1-\alpha)^2\frac{\cos(u-\alpha u)}{\sin(u-\alpha u)} - \frac{\cos u}{\sin u}\right)\geq 0
\end{align}
for which it suffices to show
\begin{align}\notag
&\alpha^2\cos\alpha u\sin(u-\alpha u)\sin u +(1-\alpha)^2\cos(u-\alpha u)\sin\alpha u\sin u-\cos u\sin\alpha u\sin(u-\alpha u)\geq 0\\\notag
\Longleftrightarrow&\alpha^2\sin^2(u-\alpha u) + (1-\alpha)^2\sin^2\alpha u - 2\alpha(1-\alpha)\cos u\sin\alpha u\sin(u-\alpha u)\geq 0\\\notag
\Longleftrightarrow&(\alpha\sin(u-\alpha u) - (1-\alpha)\sin\alpha u)^2 + 2\alpha(1-\alpha)(1-\cos u)\sin\alpha u\sin(u-\alpha u)\geq 0
\end{align}

At this point, we have proved that both $q_\alpha(u)$ and $g_\alpha(u)$ are monotonically increasing in $u\in [0,\ \pi]$ at least for $\alpha\in[0,\ 0.5]$.

\begin{align}\notag
\frac{\partial F_\alpha(t)}{\partial \alpha} = E\left(\frac{-\frac{1}{t}\frac{g_\alpha(u_2)q_\alpha(u_2)q_\alpha(u_1) - g_\alpha(u_1)q_\alpha(u_1)q_\alpha(u_2) }{q_\alpha^2(u_1)}}{\left(1+\frac{q_\alpha(u_2)}{tq_\alpha(u_1)}\right)^2}\right)
=\frac{1}{t} E\left(\frac{q_\alpha(u_1)q_\alpha(u_2)\left(g_\alpha(u_1)-g_\alpha(u_2)\right)}{\left(q_\alpha(u_1)+q_\alpha(u_2)/t\right)^2}\right)
\end{align}
By symmetry
\begin{align}\notag
\frac{\partial F_\alpha(t)}{\partial \alpha}
=\frac{1}{t} E\left(\frac{q_\alpha(u_1)q_\alpha(u_2)\left(g_\alpha(u_2)-g_\alpha(u_1)\right)}{\left(q_\alpha(u_2)+q_\alpha(u_1)/t\right)^2}\right)
\end{align}
Thus, to show $\frac{\partial F_\alpha(t)}{\partial \alpha} \geq 0$, it suffices to show
\begin{align}\notag
&E\left(\frac{q_\alpha(u_1)q_\alpha(u_2)\left(g_\alpha(u_1)-g_\alpha(u_2)\right)}{\left(q_\alpha(u_1)+q_\alpha(u_2)/t\right)^2}\right)+
E\left(\frac{q_\alpha(u_1)q_\alpha(u_2)\left(g_\alpha(u_2)-g_\alpha(u_1)\right)}{\left(q_\alpha(u_2)+q_\alpha(u_1)/t\right)^2}\right)\geq 0\\\notag
\Longleftrightarrow & E\left(\frac{q_\alpha(u_1)q_\alpha(u_2)\left(g_\alpha(u_1)-g_\alpha(u_2)\right)\left(q_\alpha^2(u_1) - q_\alpha^2(u_2)\right)\left(1/t^2-1\right)}{\left(q_\alpha(u_1)+q_\alpha(u_2)/t\right)^2\left(q_\alpha(u_2)+q_\alpha(u_1)/t\right)^2}\right)\geq 0
\end{align}
which holds because $1/t^2-1\geq 0$ and $\left(g_\alpha(u_1)-g_\alpha(u_2)\right)\left(q_\alpha(u_1) - q_\alpha(u_2)\right)\geq 0$ as both $g_\alpha(u)$ and $q_\alpha(u)$ are monotonically increasing functions of $u\in[0, \ \pi]$. This completes the proof.

\newpage

\section{Proof of Lemma~\ref{lem_F_order}}\label{app_lem_F_order}

The goal is to show that $F_\alpha(t) =\Theta\left(t^{1-\alpha}\right)$.  By our definition,
\begin{align}\notag
F_\alpha(t) = E\left(\frac{1}{1+Q_\alpha/t}\right)  =E\left(\frac{1}{1+\frac{1}{t}\frac{q_\alpha(u_2)}{q_\alpha(u_1)}}\right)
\end{align}
where
\begin{align}\notag
&q_\alpha(u) = \left[{\sin\left(\alpha u\right)}\right]^{\alpha/(1-\alpha)}\left[\frac{1}{\sin u
}\right]^{\frac{1}{1-\alpha}} {\sin\left( u - \alpha u\right)}
\end{align}

We can write the integral as
\begin{align}\notag
&F_\alpha(t) = E\left(\frac{1}{1+\frac{1}{t}\frac{q_\alpha(u_2)}{q_\alpha(u_1)}}\right)\\\notag
=&\frac{1}{\pi^2}\int_0^{\pi/2}\int_0^{\pi/2} \frac{1}{1+t^{-1}q_\alpha(u_2)/q_\alpha(u_1) }du_1du_2 + \frac{1}{\pi^2}\int_0^{\pi/2}\int_0^{\pi/2} \frac{1}{1+t^{-1}q^\prime_\alpha(u_2)/q_\alpha(u_1) }du_1du_2\\\notag
+&\frac{1}{\pi^2}\int_0^{\pi/2}\int_0^{\pi/2} \frac{1}{1+t^{-1}q_\alpha(u_2)/q^\prime_\alpha(u_1) }du_1du_2 + \frac{1}{\pi^2}\int_0^{\pi/2}\int_0^{\pi/2} \frac{1}{1+t^{-1}q^\prime_\alpha(u_2)/q^\prime_\alpha(u_1) }du_1du_2
\end{align}
where
\begin{align}\notag
q^\prime_\alpha(u) =& \left[{\sin\left(\alpha (\pi- u)\right)}\right]^{\alpha/(1-\alpha)}\left[\frac{1}{\sin(\pi- u)
}\right]^{\frac{1}{1-\alpha}} {\sin\left( \pi-u - \alpha (\pi-u)\right)}\\\notag
=&\left[{\sin\left(\alpha (\pi- u)\right)}\right]^{\alpha/(1-\alpha)}\left[\frac{1}{\sin u
}\right]^{\frac{1}{1-\alpha}} {\sin\left( u + \alpha (\pi-u)\right)}
\end{align}

First, using the fact that $ \alpha \sin u\leq \sin(\alpha u)\leq \alpha u$, we obtain
\begin{align}\notag
q_\alpha(u) \geq \left[{\alpha\sin\left( u\right)}\right]^{\alpha/(1-\alpha)}\left[\frac{1}{\sin u
}\right]^{\frac{1}{1-\alpha}} {(1-\alpha)\sin\left( u\right)} = \alpha^{\alpha/(1-\alpha)}(1-\alpha)
\end{align}
We have proved in the proof of Lemma~\ref{lem_F} that $q_\alpha(u)$ is a monotonically increasing function of $u\in[0,\ \pi]$. Since $q_\alpha(\pi/2) =  \left[{\sin\left(\alpha \pi/2\right)}\right]^{\alpha/(1-\alpha)} {\cos\left( \alpha \pi/2\right)}$, we have
\begin{align}\notag
1/4\leq \alpha^{\alpha/(1-\alpha)}(1-\alpha) \leq q_\alpha(u) \leq \left[{\sin\left(\alpha \pi/2\right)}\right]^{\alpha/(1-\alpha)} {\cos\left( \alpha \pi/2\right)}\leq 1,\hspace{0.2in} u\in [0,\ \pi/2]
\end{align}
In other words, we can view $q_\alpha(u)$ as a constant (i.e., $q_\alpha(u)\asymp1$) when $u\in[0,\ \pi/2]$.

On the other hand, note that $q^\prime_\alpha(u)\rightarrow\infty$ as $u\rightarrow 0$. Moreover, when $u\in[0,\ \pi/2]$, we have $\alpha u \leq \pi-u$ and $u-\alpha u\leq u+\alpha (\pi-u)$. Thus, $q^\prime_\alpha(u)$ dominates $q_\alpha(u)$. Therefore, the order of $F_\alpha(t)$ is determined by one term:
\begin{align}\notag
&F_\alpha(t) \asymp \int_0^{\pi/2}\int_0^{\pi/2} \frac{1}{1+t^{-1}q_\alpha(u_2)/q^\prime_\alpha(u_1) }du_1du_2 \asymp \int_0^{\pi/2} \frac{1}{1+t^{-1}/q^\prime_\alpha(u) }du
\end{align}

Since
\begin{align}\notag
q^\prime_\alpha(u) \asymp \frac{\alpha^{\alpha/(1-\alpha)}\max\{u,\alpha\}}{u^{1/(1-\alpha)}}\asymp \max\left\{u^{-\alpha/(1-\alpha)}, \alpha u^{-1/(1-\alpha)}\right\}
\end{align}
we have, for $\alpha\in[0,\ 1/2]$,
\begin{align}\notag
F_\alpha(t) \asymp& \int_0^{\alpha} \frac{1}{1+t^{-1}/q^\prime_\alpha(u) }du + \int_{\alpha}^{\pi/2} \frac{1}{1+t^{-1}/q^\prime_\alpha(u) }du\\\notag
\asymp& \int_0^{\alpha} \frac{1}{1+(\alpha t)^{-1} u^{1/(1-\alpha)}}du + \int_{\alpha}^{\pi/2} \frac{1}{1+t^{-1}u^{\alpha/(1-\alpha)} }du\\\notag
\end{align}

Consider $t<\alpha^{\alpha/(1-\alpha)}$. Because $t^{-1}u^{\alpha/(1-\alpha)} >(u/\alpha)^{\alpha/(1-\alpha)}\geq1$ for $u\geq \alpha$, we have
\begin{align}\notag
& \int_{\alpha}^{\pi/2} \frac{1}{1+t^{-1}u^{\alpha/(1-\alpha)} }du\asymp \int_{\alpha}^{\pi/2} \frac{1}{t^{-1}u^{\alpha/(1-\alpha)} }du
=t\left. \frac{1-\alpha}{1-2\alpha}u^{(1-2\alpha)/(1-\alpha)}\right|_\alpha^{\pi/2}\asymp t
\end{align}
uniformly for $\alpha<1/2$. When $\alpha=1/2$ (i.e., $t<1/2$), we also have
\begin{align}\notag
& \int_{\alpha}^{\pi/2} \frac{1}{1+t^{-1}u^{\alpha/(1-\alpha)} }du = \int_{1/2}^{\pi/2} \frac{1}{1+t^{-1}u}du = t\left.\log (u+t)\right|_{1/2}^{\pi/2}\asymp t
\end{align}
For the other term with $u\in[0,\alpha]$, we have
\begin{align}\notag
\int_0^{\alpha} \frac{1}{1+(\alpha t)^{-1} u^{1/(1-\alpha)}}du
=&\int_0^{(\alpha t)^{1-\alpha}} \frac{1}{1+(\alpha t)^{-1} u^{1/(1-\alpha)}}du+\int_{(\alpha t)^{1-\alpha}}^\alpha \frac{1}{1+(\alpha t)^{-1} u^{1/(1-\alpha)}}du\\\notag
=&\int_0^{(\alpha t)^{1-\alpha}} \frac{1}{1+(\alpha t)^{-1} u^{1/(1-\alpha)}}du+\int_{(\alpha t)^{1-\alpha}}^\alpha \frac{1}{1+(\alpha t)^{-1} u^{1/(1-\alpha)}}du\\\notag
\asymp& (\alpha t)^{1-\alpha} - (\alpha t)\frac{1-\alpha}{\alpha}\left. u^{(-\alpha)/(1-\alpha)} \right|_{(\alpha t)^{1/(1-\alpha)}}^\alpha \\\notag
=&(\alpha t)^{1-\alpha} - t(1-\alpha)\alpha^{(-\alpha)/(1-\alpha)} + t(1-\alpha)(\alpha t)^{-\alpha}\\\notag
=&t^{1-\alpha} \alpha^{-\alpha} - t(1-\alpha)\alpha^{(-\alpha)/(1-\alpha)}
\end{align}
Combining the results, we obtain
\begin{align}\notag
F_\alpha(t) \asymp&  t\left(1-\alpha^{(-\alpha)/(1-\alpha)}+\alpha^{(1-2\alpha)/(1-\alpha)}\right) + t^{1-\alpha} \alpha^{-\alpha}\asymp t^{1-\alpha}
\end{align}

This completes the proof.

\section{Proof of Lemma~\ref{lem_ratio_MLE}}\label{app_lem_ratio_MLE}

Define  $F_Z(t) = \mathbf{Pr}\left(\frac{y_j}{s_{ij}}\leq t\right)$ and $f_Z(t) = F^\prime_Z(t)$. To find the MLE of $x_i$, we need to maximize $\prod_{j=1}^M f_Z(z_{i,j})$.  Using the result in Lemma~\ref{lem_F}, for $S_1, S_2\sim S(\alpha,1,1)$, we have
\begin{align}\notag
F_Z(t) = \mathbf{Pr}\left(\frac{y_j}{s_{ij}}\leq t\right) = \mathbf{Pr}\left(S_2/S_1\leq \frac{t-x_i}{\theta_i}\right)  = E\left(\frac{1}{1+\left(\frac{\theta_i}{t-x_i}\right)^{\alpha/(1-\alpha)} Q_\alpha}\right)
\end{align}
\begin{align}\notag
f_Z(t) = E\left(\frac{\theta_i^{\alpha/(1-\alpha)}Q_\alpha\alpha/(1-\alpha)(t-x_i)^{-1/(1-\alpha)}}{\left(1+\left(\frac{\theta_i}{t-x_i}\right)^{\alpha/(1-\alpha)} Q_\alpha\right)^2}\right)
\end{align}
\begin{align}\notag
f^\prime_Z(t) = E\left(\frac{A}{\left(1+\left(\frac{\theta_i}{t-x_i}\right)^{\alpha/(1-\alpha)} Q_\alpha\right)^4}\right)
\end{align}
where $Q_\alpha$ is defined in Lemma~\ref{lem_F} and
\begin{align}\notag
A =& \theta_i^{\alpha/(1-\alpha)}Q_\alpha\alpha/(1-\alpha)(-1/(1-\alpha))(t-x_i)^{-1/(1-\alpha)-1} \left(1+\left(\frac{\theta_i}{t-x_i}\right)^{\alpha/(1-\alpha)} Q_\alpha\right)^2\\\notag
&+2\left(1+\left(\frac{\theta_i}{t-x_i}\right)^{\alpha/(1-\alpha)} Q_\alpha\right)\left(\theta_i^{\alpha/(1-\alpha)}Q_\alpha\alpha/(1-\alpha)(t-x_i)^{-1/(1-\alpha)}\right)^2\\\notag
=&\left(1+\left(\frac{\theta_i}{t-x_i}\right)^{\alpha/(1-\alpha)} Q_\alpha\right)\theta_i^{\alpha/(1-\alpha)}Q_\alpha\alpha/(1-\alpha)^2(t-x_i)^{-1/(1-\alpha)-1}\\\notag
&\times\left(-\left(1+\left(\frac{\theta_i}{t-x_i}\right)^{\alpha/(1-\alpha)} Q_\alpha\right)+2\theta_i^{\alpha/(1-\alpha)}Q_\alpha\alpha(t-x_i)^{-\alpha/(1-\alpha)}\right)\\\notag
=&\left(1+\left(\frac{\theta_i}{t-x_i}\right)^{\alpha/(1-\alpha)} Q_\alpha\right)\theta_i^{\alpha/(1-\alpha)}Q_\alpha\alpha/(1-\alpha)^2(t-x_i)^{-1/(1-\alpha)-1}\\\notag
&\times\left(-1-\left(\frac{\theta_i}{t-x_i}\right)^{\alpha/(1-\alpha)}\left(1-2\alpha\right)\right)
\end{align}
$A\leq 0$ if $\alpha\leq 0.5$. This means, $f_Z(t)\rightarrow \infty$ when $t\rightarrow x_i$ and $f_Z(t)$ is nondecreasing in $t\geq x_i$ if $\alpha\leq 0.5$. Therefore, given $M$ observations, $z_{i,j} = y_j/s_{ij}$, the MLE is  the sample minimum. This completes the proof.

\section{Proof of Lemma~\ref{lem_bias}}\label{app_lem_bias}

\begin{align}\notag
E\left(\hat{x}_{i,min}\right) =& x_i + \int_{x_i}^\infty \mathbf{Pr}\left(\hat{x}_{i,min} > t\right) dt\\\notag
=&x_i + \int_{x_i}^\infty \left[1-F_\alpha\left(\left(\frac{t-x_i}{\theta_i}\right)^{\alpha/(1-\alpha)}\right)\right]^Mdt\\\notag
=&x_i + \theta_i\int_{0}^\infty \left[1-F_\alpha\left(\left(t\right)^{\alpha/(1-\alpha)}\right)\right]^Mdt\\\notag
=&x_i+\theta_iD_{M,\alpha}
\end{align}
We have proved in Lemma~\ref{lem_F} that
\begin{align}\notag
\frac{1}{1+1/t} = F_{0}(t) \leq F_\alpha(t) \leq F_{0.5}(t) = \frac{2}{\pi}\tan^{-1}\sqrt{t}
\end{align}
Thus,
\begin{align}\notag
D_{M,\alpha}=&\int_{0}^\infty \left[1-F_\alpha\left(\left(t\right)^{\alpha/(1-\alpha)}\right)\right]^Mdt\\\notag
\leq& \int_{0}^\infty \left[\frac{1}{1+\left(t\right)^{\alpha/(1-\alpha)}}\right]^Mdt\\\notag
=&\frac{1-\alpha}{\alpha}\int_{0}^1 t^M\left(1/t-1\right)^{(1-\alpha)/\alpha-1}\frac{1}{t^2}dt\\\notag
=&\frac{1-\alpha}{\alpha}\int_{0}^1 t^{M-(1-\alpha)/\alpha-1}\left(1-t\right)^{(1-\alpha)/\alpha-1}dt\\\notag
=&\frac{1-\alpha}{\alpha}Beta\left(M-(1-\alpha)/\alpha,\ (1-\alpha)/\alpha\right)
\end{align}

When $\alpha = 0.5$, then $\alpha/(1-\alpha)=1$, and
\begin{align}\notag
D_{M,\alpha=0.5}=&\int_{0}^\infty \left[1-F_\alpha\left(\left(t\right)^{\alpha/(1-\alpha)}\right)\right]^Mdt
= \int_{0}^\infty \left[1-\frac{2}{\pi}\tan^{-1}t\right]^Mdt\\\notag
=&\int_0^{\pi/2}\left[1-\frac{2u}{\pi}\right]^M d \tan^2{u}
 =\int_0^{\pi/2}\left[1-\frac{2u}{\pi}\right]^M d \frac{1}{\cos^2u} \\\notag
 =&\int_1^{0}u^M d \frac{1}{\sin^2\left(u\pi/2\right)}
 =M\int_0^1 \frac{u^{M-1}}{\sin^2\left(u\pi/2\right)} du -1\\\notag
 =&M\left(\frac{2}{\pi}\right)^{M}\int_0^{\pi/2} \frac{u^{M-1}}{\sin^2u}du -1
\end{align}

From the integral table \cite[2.643.7]{Book:Gradshteyn_07}, we have
\begin{align}\notag
\int \frac{u^n}{\sin^2u} du = -u^n \frac{\cos u}{\sin u} + \frac{n}{n-1} u^{n-1} + n\sum_{j=1}^\infty
(-1)^j \frac{2^{2j} u^{n+2j-1}}{(n+2j-1)(2j)!} B_{2j}
\end{align}
Therefore, to facilitate numerical calculations, we resort to (let  $n=M-1$)
\begin{align}\notag
\int_0^{\pi/2} \frac{u^{M-1}}{\sin^2u}du  =& \frac{M-1}{M-2} (\pi/2)^{M-2} + (M-1)\sum_{j=1}^\infty
(-1)^j \frac{2^{2j} (\pi/2)^{M+2j-2}}{(M+2j-2)(2j)!} B_{2j}\\\notag
=&\left(\frac{\pi}{2}\right)^{M} \left(\frac{M-1}{M-2}(\pi/2)^{-2}+ (M-1)\sum_{j=1}^\infty
(-1)^j \frac{2^{2j} (\pi/2)^{2j-2}}{(M+2j-2)(2j)!} B_{2j} \right)
\end{align}
where $B_j$ is the Bernoulli number satisfying
\begin{align}\notag
\frac{t}{e^t-1} = \sum_{j=0}^\infty B_j \frac{t^j}{j!}=\sum_{j=0}^\infty B_{2j} \frac{t^{2j}}{(2j)!} - \frac{t}{2}
\end{align}
and $B_0=1$, $B_1=-1/2$, $B_2 = 1/6$, $B_4=-1/30$, $B_6=1/42$, $B_8=-1/30$, $B_{10} = 5/66$, ...

\begin{align}\notag
D_{M,\alpha=0.5}=&M \left(\frac{M-1}{M-2}(\pi/2)^{-2}+ (M-1)\sum_{j=1}^\infty
(-1)^j \frac{2^{2j} (\pi/2)^{2j-2}}{(M+2j-2)(2j)!} B_{2j} \right)-1\\\notag
=& M(M-1)\sum_{j=0}^\infty
(-1)^j \frac{2^{2j} (\pi/2)^{2j-2}}{(M+2j-2)(2j)!} B_{2j} -1\\\notag
=&M(M-1)\frac{4}{\pi^2}\sum_{j=0}^\infty
 \frac{(-1)^j\pi^{2j}}{(M+2j-2)(2j)!} B_{2j} -1
\end{align}

This completes the proof.

\end{document}